\documentclass[iop]{emulateapj}
\begin{document}

\title{Connection between the Accretion Disk and Jet in the Radio Galaxy 3C~111}

\author{Ritaban Chatterjee\altaffilmark{1}, Alan P. Marscher\altaffilmark{2}, Svetlana G. Jorstad\altaffilmark{2,3}, Alex Markowitz\altaffilmark{4}, Elizabeth Rivers\altaffilmark{4}, Richard E. Rothschild\altaffilmark{4}, Ian M. McHardy\altaffilmark{5}, Margo F. Aller \altaffilmark{6}, Hugh D. Aller\altaffilmark{6}, Anne L\"ahteenm\"aki\altaffilmark{7}, Merja Tornikoski\altaffilmark{7}, Brandon Harrison\altaffilmark{2}, Iv\'an Agudo\altaffilmark{2,8}, Jos\'e L. G\'omez\altaffilmark{8}, Brian W. Taylor\altaffilmark{2,9}, Mark Gurwell\altaffilmark{10}}

\altaffiltext{1}{Department of Astronomy, Yale University, PO Box 208101, New Haven, CT 06520-8101; ritaban.chatterjee@yale.edu}
\altaffiltext{2}{Institute for Astrophysical Research, Boston University, 725 Commonwealth Avenue, Boston, MA 02215}
\altaffiltext{3}{Astronomical Institute of St. Petersburg State University, Universitetskij Pr. 28, Petrodvorets, 198504 St. Petersburg, Russia}
\altaffiltext{4}{Center for Astrophysics and Space Sciences, University of California, San Diego, M.C. 0424, La Jolla, CA, 92093-0424, USA }
\altaffiltext{5}{Department of Physics and Astronomy, University of Southampton, Southampton, SO17 1BJ, United Kingdom}
\altaffiltext{6}{Astronomy Department, University of Michigan, 830 Dennison, 501 East University Street, Ann Arbor, Michigan 48109-1042}
\altaffiltext{7}{Aalto University, Mets\"ahovi Radio Observatory, Mets\"ahovintie 114, FIN-02540, Kylm\"al\"a, Finland}
\altaffiltext{8}{Instituto de Astrofisica de Andalucia, CSIC, Apartado 3004, 18080 Granada, Spain }
\altaffiltext{9}{Lowell Observatory, 1400 West Mars Hill Road, Flagstaff, AZ., USA}
\altaffiltext{10}{Harvard-Smithsonian Center for Astrophysics, MS 42, 60 Garden Street, Cambridge, MA 02138, USA}

\begin{abstract}
We present the results of extensive multi-frequency monitoring of the radio galaxy 3C 111 between 2004 and 2010 at X-ray (2.4--10 keV), optical ($R$ band), and radio (14.5, 37, and 230 GHz) wave bands, as well as multi-epoch imaging with the Very Long Baseline Array (VLBA) at 43 GHz. Over the six years of observation, significant dips in the X-ray light curve are followed by ejections of bright superluminal knots in the VLBA images. This shows a clear connection between the radiative state near the black hole, where the X-rays are produced, and events in the jet. The X-ray continuum flux and Fe line intensity are strongly correlated, with a time lag shorter than 90 days and consistent with zero. This implies that the Fe line is generated within 90 light-days of the source of the X-ray continuum. The power spectral density function of X-ray variations contains a break, with steeper slope at shorter timescales. The break timescale of $13^{+12}_{-6}$ days is commensurate with scaling according to the mass of the central black hole based on observations of Seyfert galaxies and black hole X-ray binaries (BHXRBs). The data are consistent with the standard paradigm, in which the X-rays are predominantly produced by inverse Compton scattering of thermal optical/UV seed photons from the accretion disk by a distribution of hot electrons --- the corona --- situated near the disk. Most of the optical emission is generated in the accretion disk due to reprocessing of the X-ray emission.  The relationships that we have uncovered between the accretion disk and the jet in 3C 111, as well as in the FR I radio galaxy 3C 120 in a previous paper,  support the paradigm that active galactic nuclei and Galactic BHXRBs are fundamentally similar, with characteristic time and size scales proportional to the mass of the central black hole.\\
\end{abstract}


\section{Introduction}
In many cosmic systems containing jets, e.g., active galactic nuclei (AGNs), stellar mass black hole X-ray binaries (BHXRBs), and young stellar objects, the presence of an accretion disk is inferred from observations. One of the prominent theories of jet production asserts that the jet plasma is propelled by magnetic field lines that thread the accretion disk (and perhaps, in AGNs and BHXRBs, the black hole) and are twisted by differential rotation \citep{bla82,lov91,beg95,mei01,vla04}. This suggests that any relations observed between events in the accretion disk and those in the jet can be used to characterize the disk-jet connection and therefore to constrain models of jet formation. In BHXRBs, the connection between accretion state and events in the jet is well-established. In these objects, transitions to high-soft X-ray states are associated with the emergence of very bright features that proceed to propagate down the radio jet \citep{fen04_2,fen09}.

Observations related to the disk-jet connection in AGNs are complicated, since a single AGN usually does not show the entire range of properties that we wish to analyze. For example, in Seyfert galaxies, most of the observed X-ray (and optical) emission is generated in the accretion disk-corona region, but their radio jets tend to be weak and non-relativistic \citep[e.g.,][]{ulv99}. On the other hand, in radio-loud AGNs with strong, highly variable nonthermal radiation (blazars), the Doppler beamed emission from the jet at most wavelengths masks the thermal emission from the accretion disk and its nearby regions. The radio galaxies \object{3C~120} and \object{3C~111} provide an excellent opportunity to study the relationship between events in the accretion disk and those in the jet. At optical and X-ray frequencies, their properties are similar to Seyfert galaxies, e.g., the presence of strong broad emission lines in the optical band \citep{sar77}, and an iron emission line in the X-ray spectrum \citep{lew05}, while at centimeter and millimeter wavelengths their emission is dominated by a bright relativistic jet. Nevertheless, the source of the optical continuum emission in radio galaxies is not well-known. For example, it may be blackbody emission from the accretion disk \citep{mal83} or synchrotron radiation from the jet.

An accretion disk-jet connection in AGNs has been suggested by \citet{mar02}. During $\sim$3 yr of monitoring of the radio galaxy 3C~120 (1997 to 2000), each of four dips in the X-ray flux, accompanied by spectral hardening, preceded the appearance of a bright knot moving down the radio jet at a superluminal apparent speed. Using similar, but more extensive, multi-frequency monitoring observations of this object between 2002 and 2007 at X-ray (2.4--10 keV) and radio (37 GHz) wave bands as well as VLBA images at 43 GHz, \citet{cha09} established an unambiguous connection between events in the accretion disk and jet. They showed that, during this interval, the X-ray and 37 GHz variations are anti-correlated, with the X-ray variations leading those in the radio by $\sim$120 days. Furthermore, nearly every X-ray dip is followed by the ejection of a new knot in the VLBA images. These findings imply that a decrease in the X-ray production is linked with increased speed in the jet flow, causing a shock wave to form and propagate downstream, appearing as a knot of bright emission in the jet. In 3C~120, the presence of an X-ray Fe emission line implies that most of the X-rays are produced in the accretion disk region. Therefore, the above pattern is strong evidence for a disk-jet connection.

In this work, we present the results of extensive multi-frequency monitoring of 3C~111 between 2004 and 2009 at X-ray energies (2.4--10 keV), optical $R$ band, and radio frequencies 14.5, 37, and 230 GHz, as well as imaging with the Very Long Baseline Array (VLBA) at 43 GHz. We use these data to investigate the presence of a disk-jet connection similar to that in 3C~120. 3C~111 is a relatively nearby ($z=0.049$) broad-line radio galaxy (BLRG) \citep{sar77,lew05}. It is classified as a Fanaroff-Riley class II (FR II) radio galaxy \citep{fan74} containing two radio lobes with hot spots and a single-sided jet \citep{lin84}. At radio frequencies, 3C~111 has blazar-like behavior, although the jet lies at an angle $\sim$$18^{\circ}$ to our line of sight \citep{jor05}, significantly greater than is the case for typical blazars.  The counter-jet is too faint to detect, presumably because of Doppler de-boosting. 3C~111 exhibits the brightest compact radio core at cm/mm wavelengths of all FR II radio galaxies and a blazar-like spectral energy distribution \citep{kad08,sgu05}. It was one of the first radio galaxies in which superluminal motion was detected \citep{got87,pre87}; knots are ejected 1-2 times per year with typical apparent speeds of $3$-$5c$ \citep{jor05}. Recently, \citet{har08} showed that part of the $\gamma$-ray emission from the EGRET source 3EG J0416+3650 is associated with 3C~111, and integration over the first 11 months of observations with the Large Area Telescope of the {\it Fermi} Space Telescope detected 3C~111 at a rather low flux level, $1.5\pm 0.5 \times 10^{-9}$ phot cm$^{-2}$ s$^{-1}$ at energies exceeding 1 GeV \citep{abdo10}. At optical and X-ray frequencies, 3C~111 possesses properties similar to Seyfert galaxies and BHXRBs. It has a prominent iron emission line at a rest energy of 6.4 keV \citep[e.g.,][]{rey98,era00}. This implies that most of the X-rays are produced in the immediate environs of the accretion disk: the corona, a hot wind, or the base of the jet \citep{haa94,mar05,nay00}.

Power spectral density (PSD) analysis is a common technique to characterize time variability, which in turn is a powerful diagnostic of the geometry and physics of the accretion disk, corona, and jet. The PSD corresponds to the power in the variability of emission as a function of timescale. \citet{mar09} and \citet{cha09} found that the X-ray PSD of 3C~120 can be fit by a piece-wise power law, with a slope that steepens above a break frequency. This property of 3C~120 is similar to Seyfert galaxies and BHXRBs, although in many cases the PSDs of the latter have more than one break, as well as sharp peaks representing quasi-periodic oscillations. The PSD break frequency in BHXRBs and Seyfert galaxies scales inversely with the mass of the black hole \citep{ede99,now99,pou01,utt02,mar03,mch04}. \citet{mch06} showed that the break frequency and BH mass are inversely related over a range of BH masses from $10$ to $10^8 M_{\sun}$, and that, for a given black hole mass, the break frequency also increases with higher accretion rate ($\dot{m}$). 

\begin{deluxetable*}{ccccccc}
\tablewidth{0pt}
\tablecaption{Parameters of the Light Curves.\label{data}}
\tablehead{
\colhead{ } &\colhead{Data set} & \colhead{Start }& \colhead{End} & \colhead{T(days) \tablenotemark{a}} & \colhead{$\Delta$T (days) \tablenotemark{b}}  & \colhead{N$_{points}$  \tablenotemark{c}}}
\startdata
        & Monitor & 2004 March & 2010 April &2246.0 &15.0 & 822\\ 
 X-ray  & Medium & 2006 November & 2007 January  &56.0  &0.25  & 224\\
        & Longlook & 2009 February 16 & 2009 February 17 & 1.2 & 0.01& 120 \\
\tableline    
 Optical & Monitor & 2004 November & 2010 August   & 2080.0 & 15.0   & 336 \\
	    & Medium & 2005 July           & 2006 January & 160.0  & 5.0 & 37 \\ 
\tableline
 14.5 GHz  & Monitor  & 2004 February & 2010 February & 2216.0 &  - & 398\\
 37 GHz  & Monitor  & 2005 January   & 2008 December  & 1458.0  &  - & 98\\
230 GHz & Monitor  & 2004 January   & 2010 April        & 2288.0  &  - & 161
\enddata     
\tablenotetext{a}{Total length of light curves.}
\tablenotetext{b}{Bin size.}
\tablenotetext{c}{Number of data points in the unbinned light curve. \\}
\end{deluxetable*}
Here, we use our data set to verify that the X-ray PSD of 3C~111 also has a break. Measuring the PSD of 3C111 gives us an opportunity to test whether another radio-loud AGN (in addition to 3C~120) has a broadband X-ray PSD similar to those of other Seyfert galaxies and BHXRBs. We can investigate whether the break timescale ($T_{\rm B}$) is consistent with the empirical relation between $T_{\rm B}$, black hole mass ($M_{BH}$), and bolometric luminosity ($L_{\rm bol}$) proposed by \citet{mch06}, based on a sample of mainly radio-quiet Seyfert galaxies. We analyze the X-ray and radio light curves as well as the sequence of VLBA images to test whether the X-ray dips and ejection of new radio knots in the pc scale jet are temporally related, as is the case in 3C 120. In order to investigate the source of optical emission in 3C~111, we cross-correlate the optical flux variations with those at X-ray energies. Uneven sampling, as invariably occurs, can cause the correlation coefficients to be artificially low. In addition, the time lags can vary across the years owing to physical changes in the source. In light of these issues, we estimate the significance of the derived correlation coefficients by repeating the analysis with simulated light curves, based on the underlying PSD. We also follow the variation in the degree of polarization of the optical emission in order to discriminate between nonthermal and thermal emission.

\begin{figure}
\epsscale{1.1}
\plotone{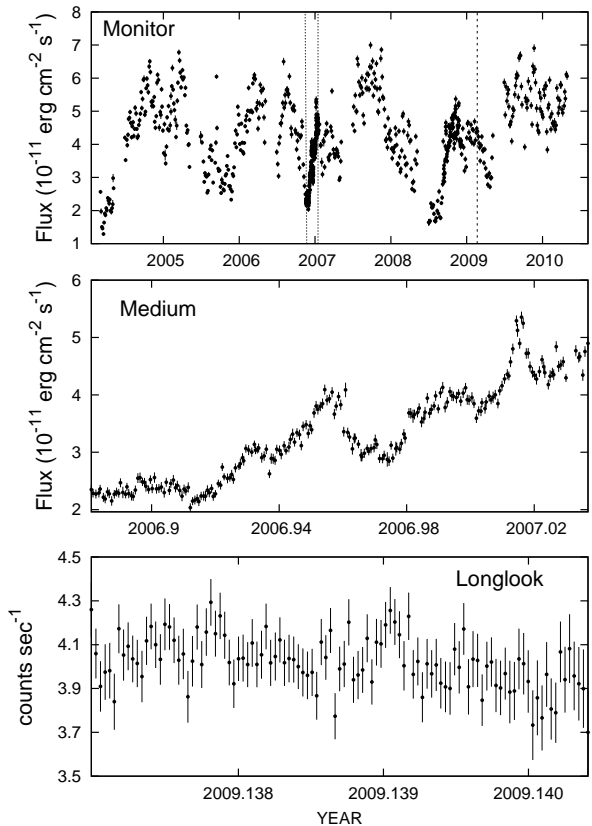}
\caption{X-ray light curves of 3C~111 with different sampling rates. The dotted and dashed vertical lines at the top panel denote the intervals shown in the middle and bottom panel, respectively. \\}
\label{xdata}
\end{figure}

In {\S}2 we present the observations and data reduction procedures, while in {\S}3 we describe the correlated variations in the X-ray continuum and Fe K$\alpha$ emission line. In {\S}4 we carry out the power spectral analysis and discuss its results and implications. In {\S}5 we investigate the relation between the X-ray dips and ejections of superluminal radio knots, as well as the X-ray/radio flux correlation. In {\S}6 we cross-correlate the X-ray and optical variations and discuss the possible sources of optical emission. In {\S}7 we compare the properties of 3C~111 with those of the radio galaxy 3C~120, while {\S}8 contains the summary and conclusions.

\section{Observations and Data Analysis}
Table~\ref{data} summarizes the intervals of monitoring at different frequencies for each of the three wave bands in our data set. We term the entire light curve ``monitor data''; shorter segments of more intense monitoring are described below.
\subsection{X-Ray Monitoring}
The X-ray light curves are based on observations of 3C~111 with the PCA detector of the {\it Rossi X-ray Timing Explorer (RXTE)} from 2004 March to 2010 April, with typical exposure times of 1-2 ks. For each exposure, we used routines from the X-ray data analysis software FTOOLS and the program XSPEC version 11.6 to calculate and subtract the standard X-ray background model for the faint sources from the data and to fit the X-ray spectrum from 2.4 to 10 keV as a power law with low-energy photoelectric absorption by intervening gas in our Galaxy. For the latter, we used a hydrogen column density of $9.0 \times 10^{21}$ atoms cm$^{-2}$. This value of the absorbing column was determined by \citet{sam99} by analyzing the X-ray spectrum of 3C 111 with data from the \textit{Advanced Satellite for Cosmology and Astrophysics (ASCA)}.  Using observations from \textit{XMM-Newton}, \citet{lew05} have also obtained a value close to this. It should be noted that a giant molecular cloud is present in the foreground of 3C 111 \citep{MMB93,MM95} and it contributes significantly to the optical to X-ray extinction of emission received from 3C 111.

\begin{figure}
\epsscale{1.1}
\plotone{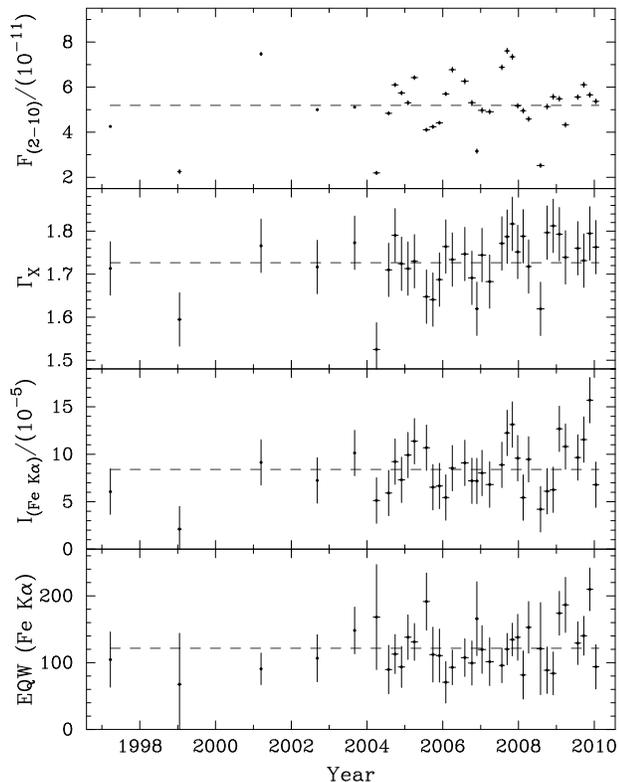}
\caption{Light curves for 2.4--10 keV continuum flux $F_{2-10}$ (in erg cm$^{-2}$ s$^{-1}$), photon index $\Gamma_{\rm X}$ (of the continuum), intensity of the Fe line $I_{\rm FeK\alpha}$ (in ph cm$^{-2}$ s$^{-1}$), and the Fe line equivalent width $EQW$ (in eV) derived from time-resolved spectral fits to the \textit{RXTE}-PCA monitoring data. The dashed lines are the mean values. \\}
\label{line_cont_lc}
\end{figure}

The sampling of the X-ray flux varied. Normally, observations were made 2-3 times per week except during 8-week intervals each year when the radio galaxy is too close to the Sun's celestial position to observe safely. In order to sample shorter-term variations, between 2006 November and 2007 January we obtained, on average, four measurements per day for almost two months. We refer to these observations as the ``medium'' data. Our monitoring program also contains an interval from 2008 September to 2008 December when we obtained roughly one data point per day. In order to sample variations on as short timescales as possible, we observed 3C~111 with the European Photon Imaging Camera (EPIC) on board \textit{XMM-Newton} continuously for about 130 ks on 2009 February 16-17. The EPIC-pn and EPIC-MOS data were obtained in the Small Window and Partial Window mode, respectively, using the thin filter. The data were processed with the the Science Analysis Software version 8.0.0. We filtered the pn data to include only single and double-pixel events (i.e., PATTERN $\leq $ 4). A pn light curve was extracted in the 2.4--10 keV energy band, similar to that of our \textit{RXTE} data, and was background-subtracted and binned to 1000 s time intervals. In general, pn data are more suitable for timing analysis and the average count rate in the pn light curve was higher than that in the MOS light curves. Hence, we use the pn light curve as the ``longlook" data. Figure~\ref{xdata} presents these three data sets. The X-ray data are given in Tables~\ref{xdatatable} and  \ref{longlookdatatable} of the electronic version.

\begin{deluxetable*}{cccccccc}
\tablewidth{0pt}
\tablecaption{X-ray (2.4-10 KeV; RXTE-PCA) light curve of 3C 111 from 2004 to 2010 (first 5 rows). \label{xdatatable}}
\tablehead{
\colhead{RJD\tablenotemark{a}} & \colhead{exp (s)} & \colhead{flux\tablenotemark{b}} & \colhead{err} & \colhead{$\alpha$\tablenotemark{c}} & \colhead{err} & \colhead{counts/s/PCU} & \colhead{err}}
\startdata
3065.9070   &   2080.   &   2.565E-11   &   5.683E-13   &   $-$0.691   &   0.062   &   2.613E+00   &   5.307E-02   \\         
3068.8816   &    752.   &   1.975E-11   &   8.709E-13   &   $-$0.530   &   0.123   &   2.035E+00   &   8.308E-02   \\         
3072.8645   &   1952.   &   1.503E-11   &   5.269E-13   &   $-$0.672   &   0.096   &   1.544E+00   &   4.893E-02   \\         
3075.8169   &   2032.   &   1.458E-11   &   5.173E-13   &   $-$0.657   &   0.100   &   1.480E+00   &   4.770E-02   \\         
3080.0244   &   1936.   &   1.292E-11   &   5.291E-13   &   $-$0.394   &   0.112   &   1.329E+00   &   4.920E-02   \\  
	...	&	...	&	...	&	...	&	...	&	...	&	...	&	...		     
\enddata
\tablenotetext{a}{Relative Julian Date$\colon$ JD$-$2450000.}
\tablenotetext{b}{Unit$\colon$ ergs cm$^{-2}$s$^{-1}$.}
\tablenotetext{c}{$\alpha\colon$ Energy Index. \\}
\end{deluxetable*}

\begin{deluxetable}{ccc}
\tablewidth{0pt}
\tablecaption{X-Ray Longlook light curve of 3C 111 from \textit{XMM-Newton} in 2.4--10 keV band on 2009 Feb 16--17 (first 5 rows). \label{longlookdatatable}}
\tablehead{
\colhead{RJD\tablenotemark{a}} & \colhead{Counts/s} & \colhead{error}}
\startdata
  4878.50   &    4.26   &   0.12   \\
  4878.51   &    4.06   &   0.11   \\
  4878.52   &    3.91   &   0.12   \\
  4878.53   &    3.98   &   0.13   \\
  4878.55   &    3.98   &   0.13   \\
 ...     & ...     & ...     
\enddata
\tablenotetext{a}{Relative Julian Date: JD$-$2450000. \\}
\end{deluxetable}

We performed time-resolved spectroscopy on the \textit{RXTE} monitoring data in order to explore the variability of the Fe K$\alpha$ emission line. If the line does indeed respond to X-ray continuum variations, then one can place constraints on the distance between the X-ray continuum source and the bulk of the line-emitting gas. The analysis closely followed that employed by previous studies \citep[e.g.,][]{vau01,mark03}. Summed spectra were constructed by combining the data from the individual $\sim$1~ks observations within the defined time bins. A time-averaged model was applied to all binned spectra. The first five time bins were defined to match the start/stop times of the \textit{RXTE} campaigns in 1997 March, 1999 January--February, 2001 March, 2001 September and 2003 September. For the sustained long-term monitoring that started on 2004 March 1, we considered all data acquired up to 2010 February 14. Bin sizes were chosen to optimize the trade-off between minimizing the uncertainties in the Fe line flux and maximizing the number of bins. This yielded 31 bins each of roughly 60 days in duration. The response of the PCA and the conversion factor between count rate per PCU and incident flux, evolve slowly. In the 2--10 keV band, 1.0 ct/s/PCU corresponds to roughly 1.12$\times10^{-11}$, 1.09$\times10^{-11}$, and 1.04$\times10^{-11}$ erg cm$^{-2}$ s$^{-1}$ during 1997, 2003, and 2008, respectively. Consequently, regarding long-term PCA light curves, fluxes obtained from spectral fitting are more valuable than count rates. We have taken this long-term trend in response into account in our analysis. The number of PCUs used during the analysis of data from various epochs are as follows: PCU 0, 1 and 2 for 1997 March, PCU 0 and 2 for 1999 Jan--Feb, since PCU 1 had started to have thermal breakdown in 1998, and PCU 2 only for all data from 2001 to 2010, since PCU 0 lost its propane layer in 2000. The number of counts in a typical binned spectrum ranged from $\sim$10$^5$ to 10$^6$. 

\begin{deluxetable*}{ccccccccc}
\tablewidth{0pt}
\tablecaption{Variation of 2.4--10 keV continuum flux, intensity of the Fe line, photon index, and the Fe line equivalent width derived from time-resolved spectral fits to the \textit{RXTE}-PCA monitoring data shown in Figure \ref{line_cont_lc}. \label{linecontdatatable}}
\tablehead{
\colhead{Yr} & \colhead{f$_{\rm cont}$\tablenotemark{a}} & \colhead{err} & \colhead{f$_{\rm line}$\tablenotemark{b}} & \colhead{err} & \colhead{$\Gamma$\tablenotemark{c}} & \colhead{err} & \colhead{EQW\tablenotemark{d}} & \colhead{err}}
\startdata
1997.22   &    4.26   &    0.03   &    6.05   &    2.39   &    1.71   &    0.06   &   104.68   &    41.40   \\
1999.05   &    2.25   &    0.09   &    2.12   &    2.39   &    1.59   &    0.06   &    67.64   &    76.44   \\
2001.20   &    7.47   &    0.07   &    9.14   &    2.39   &    1.77   &    0.06   &    90.74   &    23.76   \\
2002.69   &    5.00   &    0.04   &    7.24   &    2.39   &    1.72   &    0.06   &   106.76   &    35.29   \\
2003.67   &    5.11   &    0.03   &   10.13   &    2.39   &    1.77   &    0.06   &   148.30   &    35.03   \\
 ...     	  & ...          & ...          &       ...     &      ...     &      ...     &      ...     &       ...     &        ...      
\enddata
\tablenotetext{a}{2--10 keV continuum flux $ F_{2-10}$ ($10^{-11}$ erg cm$^{-2}$ s$^{-1}$).}
\tablenotetext{b}{Intensity of the Fe line $I_{\rm FeK\alpha}$ ($10^{-5}$ph cm$^{-2}$ s$^{-1}$).}
\tablenotetext{c}{photon index $\Gamma_{\rm X}$ of the continuum.}
\tablenotetext{d}{Fe line equivalent width $EQW$ (eV). \\}
\end{deluxetable*} 

We summed the PCA spectra from the individual observations within each time interval to create the binned spectra. PCA response matrices were generated for each individual observation using \textsc{pcarmf} v.11.7 and summed, weighted by exposure time, to yield a response for each binned spectrum. We did not include HEXTE data in the time-resolved fits, as those data yielded no additional model constraints compared to the PCA alone on these time scales given the flux level of 3C~111. Spectra were fit over the 3--25 keV energy range in \textsc{XSPEC} v.12.5.1k using a model of the same form as the best-fit to the long-term time-averaged spectrum determined by \citet{riv11}. Those authors performed joint \textit{RXTE} PCA + HEXTE spectral fitting over the 3--100 keV energy band to a summed spectrum comprising all \textit{RXTE} data on 3C~111 observed from 1997 March until 2008 July. The model we use thus consists of a power law with a photon index $\Gamma_{\rm X}$ allowed to vary after being initially set at 1.67, plus a Gaussian component to model the Fe K$\alpha$ emission line, with photo-electric absorption in our Galaxy similar to that given above. The intensity of the Gaussian component $I_{\rm FeK\alpha}$ was allowed to vary; the energy centroid and width $\sigma$ were kept fixed at 6.19 keV (rest-frame) and 0.72 keV, respectively, since there was no significant evidence to allow these parameters to vary from these best-fit time-averaged values. \citet{riv11} found no significant evidence to warrant inclusion of a Compton hump in the model. Background level corrections were applied using \textsc{recorn} within \textsc{XSPEC}.

Figure \ref{line_cont_lc} shows the resulting light curves for the 2--10 keV continuum flux $F_{2-10}$, $\Gamma_{\rm X}$, $I_{\rm FeK\alpha}$, and the Fe line equivalent width $EQW$. These data are given in Table~\ref{linecontdatatable} of the electronic version. Errors on  $I_{\rm FeK\alpha}$, $\Gamma_{\rm X}$, and $EQW$ were derived using the point-to-point variance\footnote{See $\S$3.3 of \citet{vau01} or $\S$3.3 of \citet{mar03} for further details and the definition of the point-to-point variance.}. We determined errors on values of $F_{2-10}$ within a time bin from the 2--10~keV continuum light curve, using the mean flux error on the $\sim$1~ks exposures in that time bin. 

\subsection{Optical Monitoring}
\begin{figure}
\epsscale{1.1}
\plotone{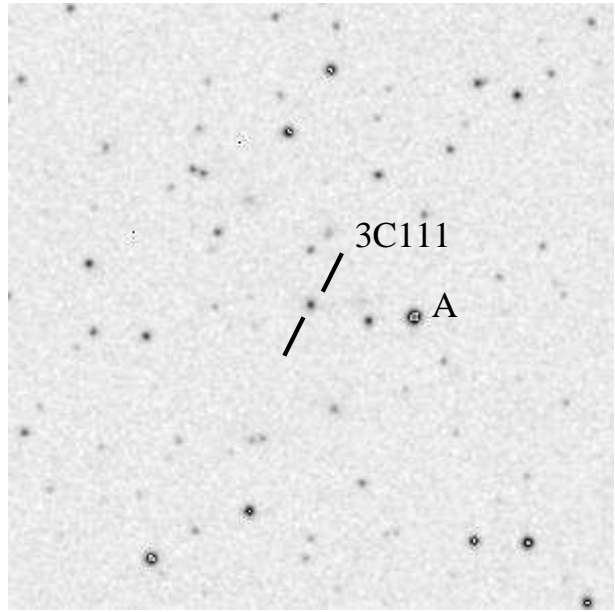}  
\caption{Finding chart for 3C 111. Star A is the comparison star with $R=14.93\pm0.05$, located $\sim$50$\arcsec$ west of 3C 111, used to calculate the magnitude and interstellar polarization correction ($p=3.6\pm0.2\%$, $\chi=123^\circ\pm2^\circ$). \\}
\label{compstar}
\end{figure}

We monitored 3C~111 in the optical $R$ band over a portion of the time span of the X-ray observations. The majority of the measurements in $R$ band are from the 2 m Liverpool Telescope (LT) at La Palma, Canary Islands, Spain, supplemented by observations at the 1.8 m Perkins Telescope of Lowell Observatory, Flagstaff, Arizona. We used star A (Figure \ref{compstar}) located $\sim$50$\arcsec$ west of 3C~111 as a comparison star, for which we determined an $R$-band magnitude $R=14.93\pm0.05$ based on differential photometry with standard stars from other fields of view. Data acquisition and analysis procedures were identical to those described by \citet{cha09}. We also monitored the $R$-band linear polarization of 3C~111 from 2006 to 2009 with the PRISM camera on the Perkins telescope at Lowell Observatory and with the 2.2 m telescope at Calar Alto Observatory, Spain\footnote{Data acquired as part of the MAPCAT program: http://www.iaa.es/$\sim$iagudo/research/MAPCAT}. Interstellar polarization (ISP) is significant in the field of 3C~111. We used the polarization parameters of the comparison star, presumed to be intrinsically unpolarized, to correct the polarization of 3C~111 for ISP \citep[see also][]{jor07}. We corrected the degree of linear polarization for statistical bias using the method of \citet{war74}.

\begin{deluxetable}{ccccc}
\tablewidth{0pt}
\tablecaption{Optical (R Band) light curve of 3C 111 from 2004 to 2010 (first 5 rows).\label{opdatatable}}
\tablehead{
\colhead{RJD\tablenotemark{a}} & \colhead{R Mag} & \colhead{err} & \colhead{Flux Density\tablenotemark{b}} & \colhead{err}}
\startdata
3325.5154     &        17.033     &         0.009     &         0.474     &         0.004     \\         
3325.5171     &        17.090     &         0.009     &         0.450     &         0.004     \\         
3325.5210     &        17.047     &         0.009     &         0.468     &         0.004     \\         
3325.5247     &        17.055     &         0.009     &         0.465     &         0.004     \\         
3325.5266     &        17.030     &         0.009     &         0.475     &         0.004     \\      
	...	&	...	&	...	&	...	&	...	
\enddata
\tablenotetext{a}{Relative Julian Date$\colon$ JD$-$2450000.}
\tablenotetext{b}{Unit$\colon$ mJy. \\}
\end{deluxetable} 

\subsection{Radio Monitoring}
We monitored the 14.5 GHz flux of 3C~111 with the 26 m antenna of the University of Michigan Radio Astronomy Observatory. Details of the calibration and analysis techniques are described in \citet{all85}. The flux scale is set by observations of \object{Cassiopeia~A} \citep{baa77}. We have compiled a 37 GHz light curve with data from the 13.7 m telescope at Mets\"ahovi Radio Observatory, Finland. The flux density scale is based on observations of \object{DR~21}, with \object{3C~84} and \object{3C~274} used as secondary calibrators. A detailed description of the data reduction and analysis is given in \citet{ter98}. We have also compiled a 230 GHz light curve of 3C~111 over a portion of the time span of the X-ray observations with calibration data from the the Submillimeter Array \citep[SMA; see][for details]{Gur07}.

\begin{deluxetable}{ccc}
\tablewidth{0pt}
\tablecaption{Radio (14.5 GHz) light curve of 3C 111 from 2004 to 2010 (first 5 rows). \label{15ghzdatatable}}
\tablehead{
\colhead{RJD\tablenotemark{a}} & \colhead{Flux Density (Jy)} & \colhead{error}}
\startdata
   3024.5   &   2.97   &   0.10   \\         
   3042.5   &   3.06   &   0.04   \\         
   3050.5   &   3.16   &   0.04   \\         
   3052.5   &   3.09   &   0.04   \\         
   3064.5   &   3.02   &   0.03   \\         
 ...     & ...     & ...    
\enddata
\tablenotetext{a}{Relative Julian Date$\colon$ JD$-$2450000. \\}
\end{deluxetable}

\begin{deluxetable}{ccc}
\tablewidth{0pt}
\tablecaption{Radio (37 GHz) light curve of 3C 111 from 2005 to 2008 (first 5 rows). \label{37ghzdatatable}}
\tablehead{
\colhead{RJD\tablenotemark{a}} & \colhead{Flux Density (Jy)} & \colhead{error}}
\startdata
 3371.22   &    4.61   &   0.25   \\         
  3372.20   &    4.35   &   0.24   \\         
  3379.52   &    4.19   &   0.22   \\         
  3423.03   &    4.23   &   0.23   \\         
  3447.93   &    4.17   &   0.26   \\         
 ...     & ...     & ...     
\enddata
\tablenotetext{a}{Relative Julian Date$\colon$ JD$-$2450000. \\}
\end{deluxetable}

\begin{deluxetable}{ccc}
\tablewidth{0pt}
\tablecaption{Radio (230 GHz) light curve of 3C 111 from 2004 to 2010 (first 5 rows). \label{230ghzdatatable}}
\tablehead{
\colhead{RJD\tablenotemark{a}} & \colhead{Flux Density (Jy)} & \colhead{error}}
\startdata
 3022.48   &    1.26   &   0.09   \\         
  3031.60   &    1.71   &   0.35   \\         
  3266.67   &    2.06   &   0.13   \\         
  3359.38   &    2.49   &   0.13   \\         
  3381.64   &    3.80   &   0.20   \\         
 ...     & ...     & ...    
\enddata
\tablenotetext{a}{Relative Julian Date$\colon$ JD$-$2450000. \\}
\end{deluxetable}

\begin{figure*}[t]
\epsscale{.60}
\plotone{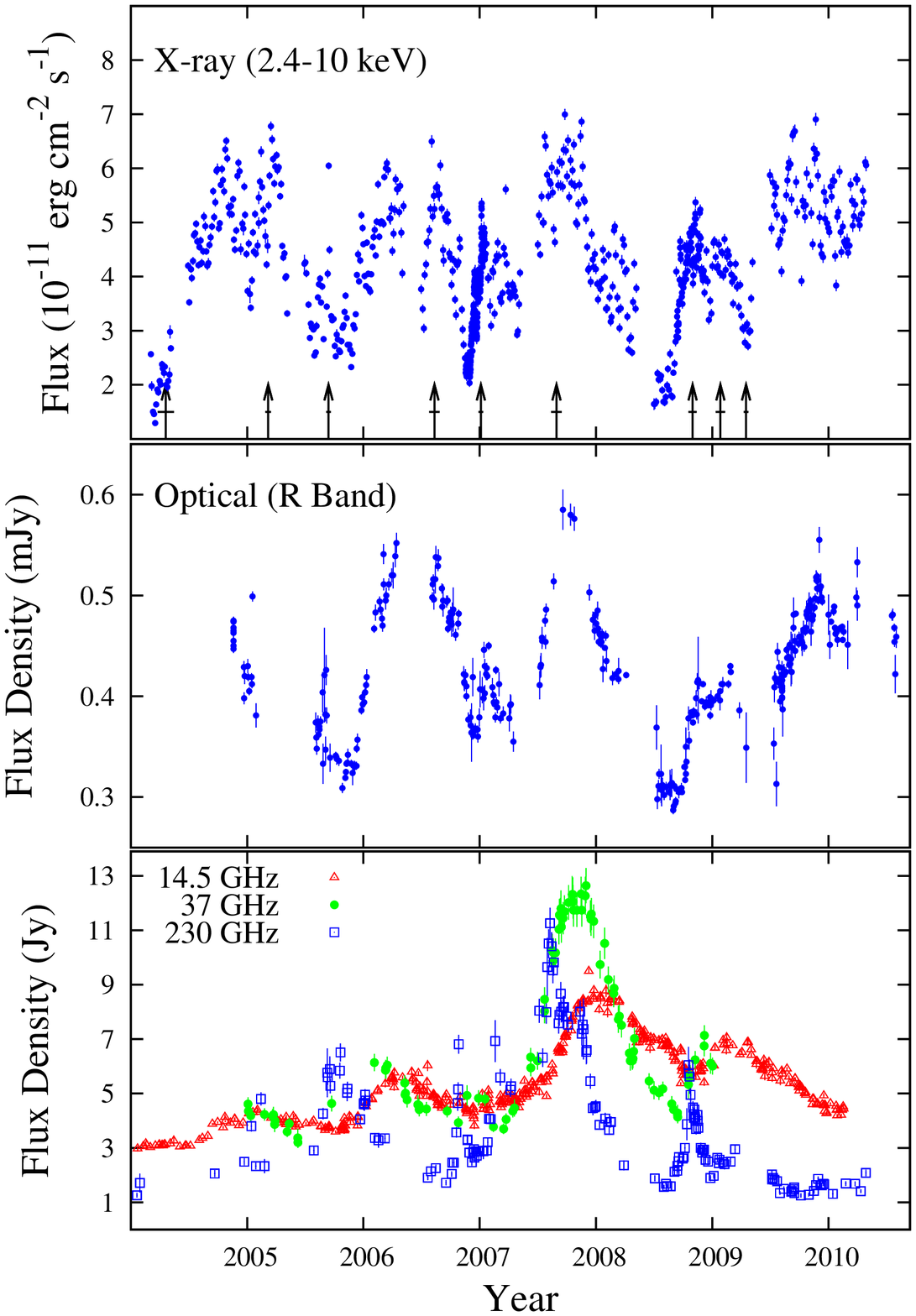} 
\caption{Variation of X-ray flux, optical flux density, and radio flux density of 3C~111 from 2004 to 2010. In the top panel, the arrows indicate the times of superluminal ejections and the line segments perpendicular to the arrows represent the uncertainties in the times. \\}
\label{xoprad}
\end{figure*}

\begin{figure}
\epsscale{1.1}
\plotone{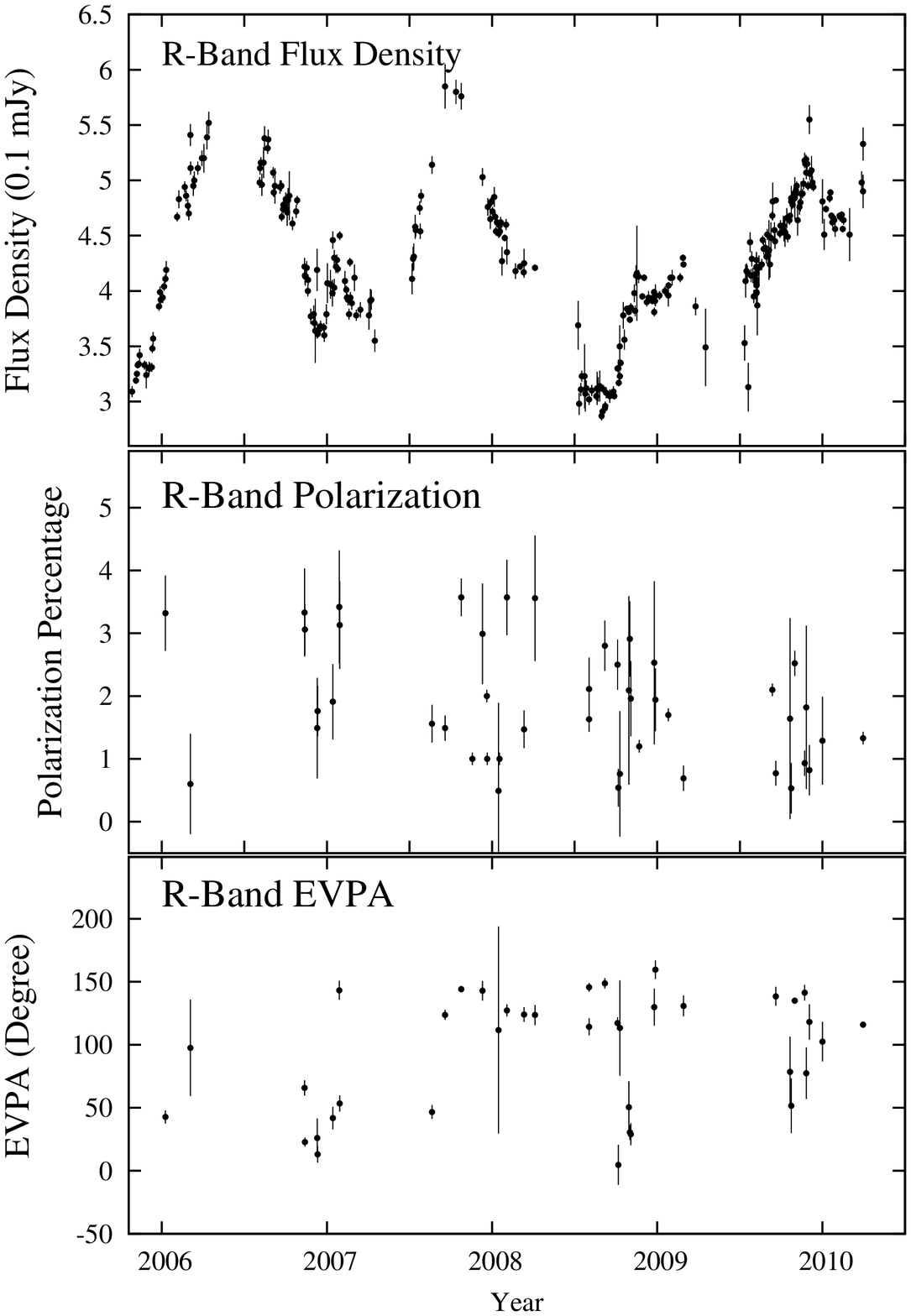}
\caption{Variation of the optical R-band flux density, polarization percentage, and electric vector position angle (EVPA) of 3C 111 during 2006 to 2010. The average polarization was only $1.6\%\pm0.6\%$, rather low for synchrotron radiation from the jet. This supports the notion that most of the optical emission is thermal radiation from the accretion disk. \\}
\label{op_pol}
\end{figure}

Figure~\ref{xoprad} presents the X-ray, optical, and radio light curves. The variation of the total R-band flux, polarization percentage, and electric vector position angle (EVPA) are shown in Fig. \ref{op_pol}. These data are given in Tables~\ref{opdatatable}, \ref{15ghzdatatable}, \ref{37ghzdatatable}, \ref{230ghzdatatable}, and \ref{poldatatable} of the electronic version.

\begin{deluxetable}{ccccc}
\tablewidth{0pt}
\tablecaption{Variation of the R-band Polarization Percentage and EVPA of 3C 111 from 2006 to 2010.\label{poldatatable}}
\tablehead{
\colhead{RJD\tablenotemark{a}} & \colhead{p (\%)\tablenotemark{b}} & \colhead{error} & \colhead{EVPA\tablenotemark{c}} & \colhead{error}}
\startdata
3741.74   &     3.32   &     0.60   &    42.70   &     5.20   \\
3796.65   &     0.60   &     0.80   &    97.50   &    38.30   \\
4048.87   &     3.33   &     0.70   &    65.80   &     6.00   \\
4049.98   &     3.06   &     0.40   &    22.70   &     3.70   \\
4076.88   &     1.49   &     0.80   &    26.00   &    15.40   \\
...   &     ...   &     ...   &    ...   &    ...  
\enddata
\tablenotetext{a}{Relative Julian Date$\colon$ JD$-$2450000.}
\tablenotetext{b}{R-band polarization percentage.}
\tablenotetext{c}{Electric vector position angle. \\}
\end{deluxetable} 

\subsection{VLBA Imaging}
\begin{figure*}
\epsscale{.8}
\plotone{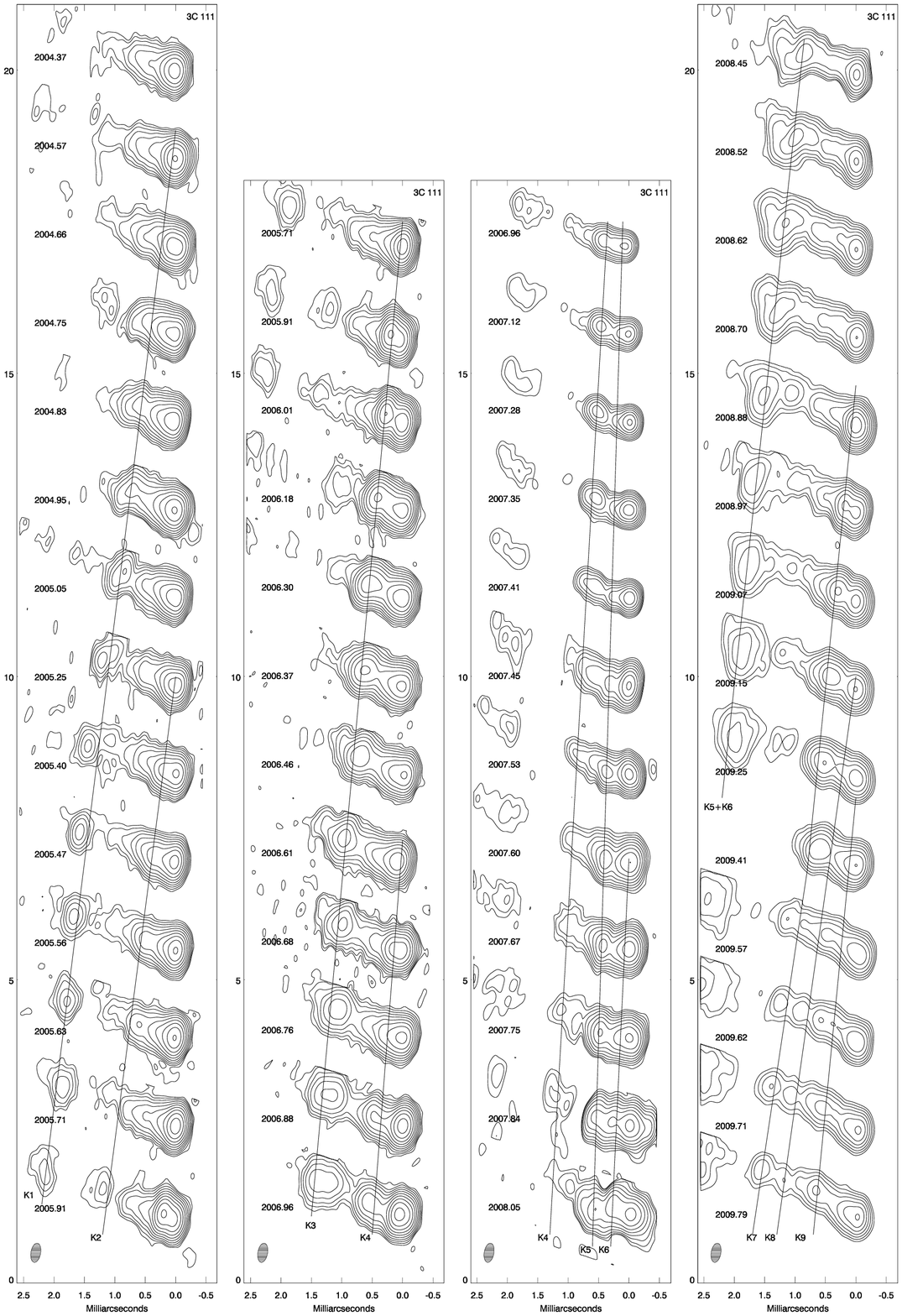}
\caption{Sequences of VLBA images at 7 mm during 2004 to 2009. The images are convolved with an elliptical Gaussian beam of FWHM size 0.32$\times$0.16 mas at PA = $-10\degr$. The global peak over all maps is 4.80 Jy/Beam. The contour levels are 0.25, 0.354, 0.5, 0.707, ..., 90.51 \% of the global peak. Note that the time spacing of the images is not uniform, hence the lines denoting the proper motion of a given knot do not pass through the centroid of the knot at every epoch. \\}
\label{vlbaimages}
\end{figure*}

\begin{figure}
\epsscale{1.1}
\plotone{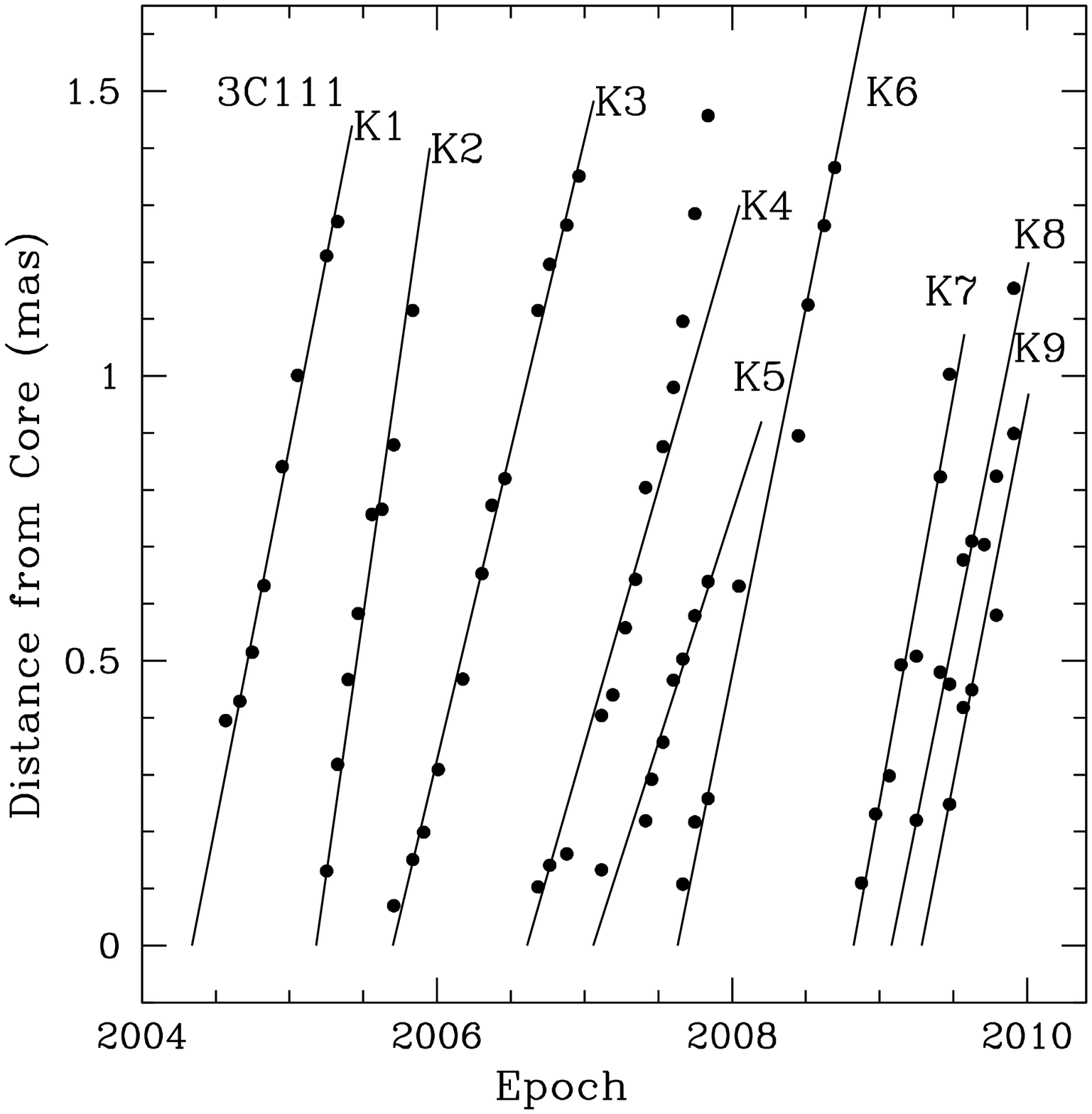}
\caption{Angular separation from the core vs. epoch of the moving knots brighter than 100 mJy within 2.0 mas of the core. The black lines indicate the motion of each knot listed in Table \ref{ejecflare}. A knot is identified through continuity of the trajectory from one epoch to the next. The identification of the knot across epochs considers its position angle relative to the core and, in some cases, its polarization. At some epochs, the model fit breaks a knot into two pieces. \\}
\label{distepoch}
\end{figure}

Starting in 2004 May, we observed 3C~111 with the Very Long Baseline Array (VLBA) at 43 GHz at roughly monthly intervals, with some gaps of 2-4 months. The sequence of images (Figure \ref{vlbaimages}) from these data provide a dynamic view of the jet at an angular resolution $\sim$0.1 milliarcseconds (mas) in the direction of the jet, corresponding to 0.094 pc for an adopted Hubble constant of $H_{0}$ = 70 km s$^{-1}$ Mpc$^{-1}$. We processed the data in the same manner as described in \citet{jor05}. We modeled the brightness distribution at each epoch with multiple components with circular Gaussian brightness distributions using the task MODELFIT of the software package Difmap \citep{she97}. This represents the jet emission downstream of the core, the bright feature situated at the western end of the jet, by a sequence of knots (also referred to as ``components''), each characterized by its flux density, FWHM diameter, and position relative to the core. The ejection time $T_0$ is the extrapolated time of coincidence of a moving knot with the position of the (presumed stationary) core. Figure~\ref{distepoch} plots the distance vs. epoch for all moving components brighter than 100 mJy within 2.0 mas of the core. We use the position vs. time data to determine the projected direction on the sky of the inner jet, as well as the apparent speeds and ejection times of new superluminal knots. Table~\ref{ejecflare} lists the values of $T_0$ and apparent speeds of new knots determined by the above procedure. In the top panel of Figure~\ref{xoprad}, the arrows represent the times of superluminal ejections, while the line segments perpendicular to the arrows show the uncertainties in the values of $T_0$. The apparent speeds of the components with well-determined motions are between 2.2 and 6.4$c$. 

\begin{deluxetable}{cccccc}
\tablewidth{0pt}
\tablecaption{Times of X-ray Dips, Ejection Times ($T_0$) of Radio Knots, and Apparent Speeds ($\beta_{\rm app}$) of Radio Knots.\label{ejecflare}}
\tablehead{
\colhead{Dip ID} & \colhead{$\rm T_{\rm Xmin}$\tablenotemark{a}} & \colhead{Knot ID} & \colhead{$\rm T_0$} & \colhead{$\beta_{\rm app}$}}
\startdata
d1 & 2004.19  & K1 &  2004.30$\pm$ 0.14 & 3.99 $\pm$ 0.19  \\
d2 & 2005.02  & K2 &  2005.18$\pm$ 0.05 & 6.37 $\pm$ 0.29 \\
d3 & 2005.57  & K3 &  2005.70$\pm$ 0.04 &  3.44 $\pm$ 0.08\\
d4 & 2006.47  & K4 &  2006.61$\pm$ 0.09 &  2.24 $\pm$ 0.44 \\
d5 & 2006.88  & K5 &  2007.01$\pm$ 0.04 &   2.71 $\pm$ 0.38 \\
d6 & 2007.32  & K6 &  2007.66$\pm$ 0.09 &  4.18 $\pm$ 0.11 \\
d7 & 2008.51  & K7 &  2008.83$\pm$ 0.07  &  4.54 $\pm$ 0.38\\
d8 & 2008.98  &  K8 & 2009.07$\pm$ 0.08  &  4.07 $\pm$ 0.43\\
d9 & 2009.26  &  K9 & 2009.29$\pm$ 0.04  &  4.33 $\pm$ 0.66
\enddata
\tablenotetext{a}{Time of X-ray minimum. \\}
\end{deluxetable}

\section{Correlated Variations in the X-ray Continuum and Fe K$\alpha$ Emission Line}
\begin{figure}
\epsscale{0.9}
\plotone{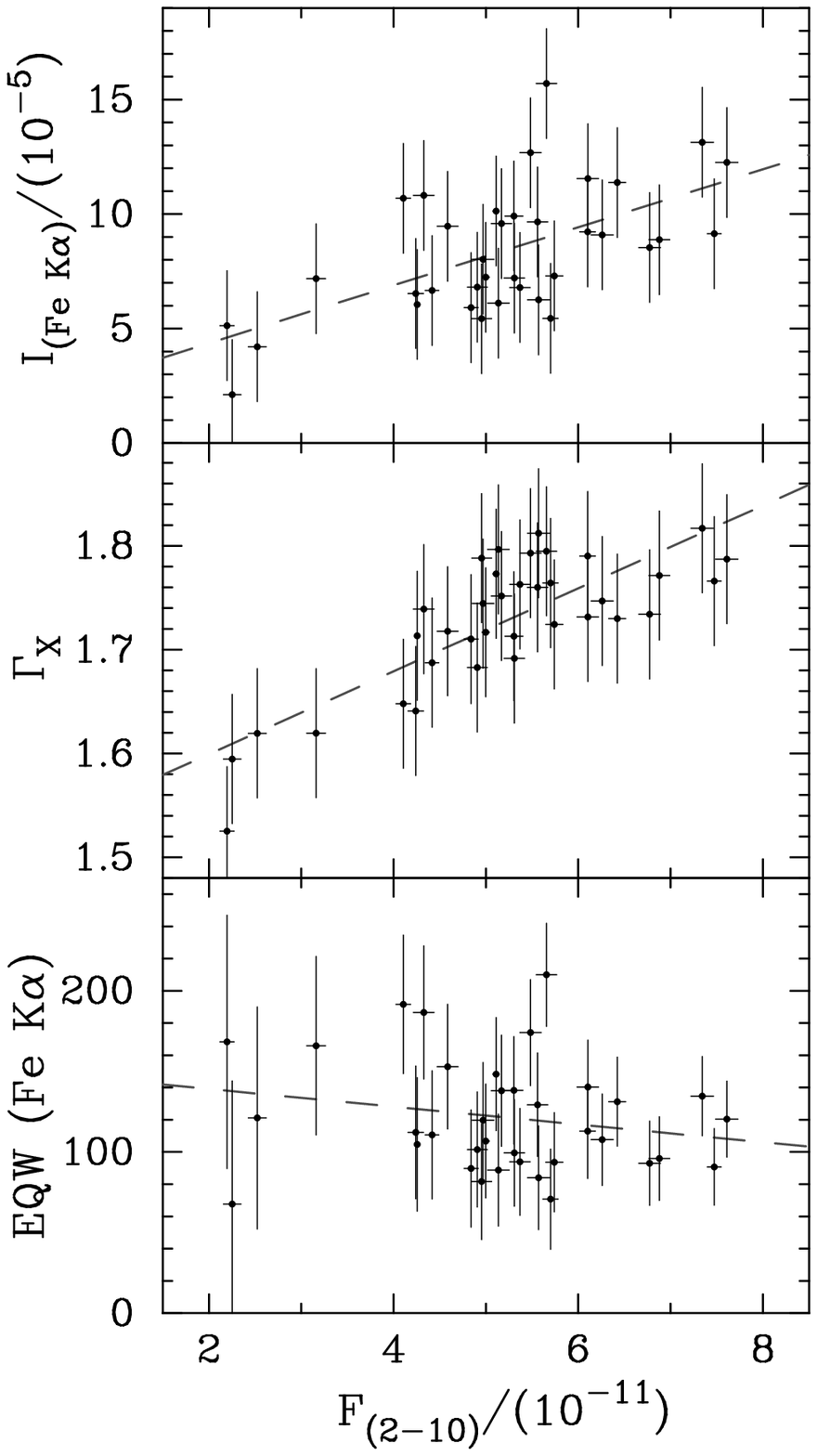}
\caption{Zero-lag correlation diagrams for intensity of the Fe line $I_{\rm FeK\alpha}$, photon index $\Gamma_{\rm X}$ (of the continuum), and Fe line equivalent width $EQW$ each plotted against 2--10 keV continuum flux $F_{2-10}$ with the best-fit linear relation plotted for each as a dashed line; units are the same as in Figure~\ref{line_cont_lc}. The X-ray continuum/Fe line correlation presented here is one of the strongest such correlation observed so far. The best-fit linear relations plotted here are summarized in Table \ref{line_cont_tab} along with the Pearson correlation coefficients. \\}
\label{line_cont_cor}
\end{figure}

\begin{figure}
\epsscale{1.1}
\plotone{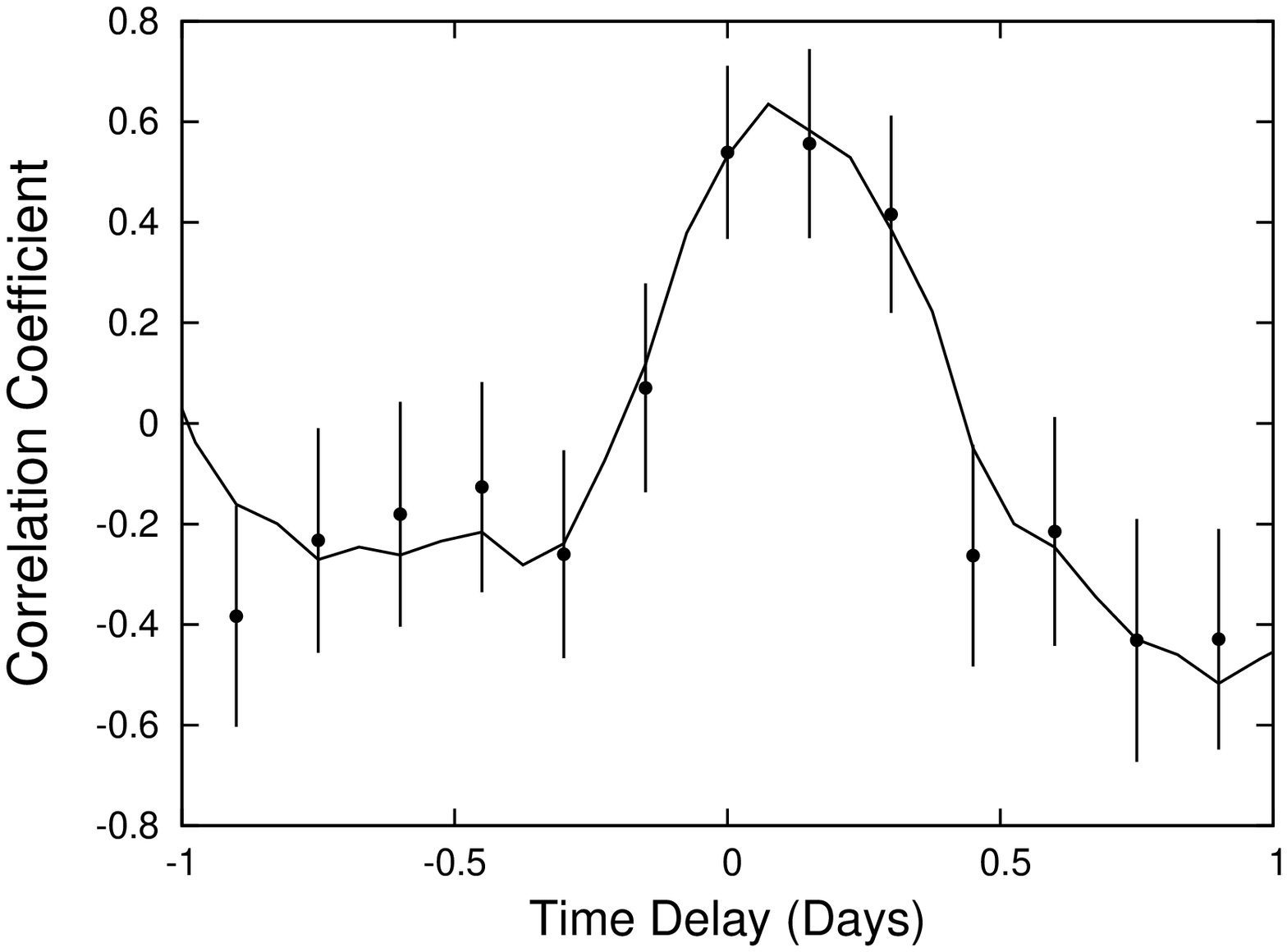}
\caption{Points denote discrete cross-correlation function (DCCF), and the solid line shows the interpolated cross-correlation function (ICCF) of the variation of intensity of the Fe line $I_{\rm FeK\alpha}$ with that of 2--10 keV continuum flux $F_{2-10}$. The two variations are strongly correlated with a time delay consistent with zero with a 1$\sigma$ upper limit of 88 days. \\}
\label{line_cont_corfunc}
\end{figure}

Figure \ref{line_cont_cor} shows the zero-lag correlation diagrams of  $I_{\rm FeK\alpha}$, $\Gamma_{\rm X}$, and $EQW$, each plotted against $F_{2-10}$. Best-fit linear relations are plotted for each and are summarized in Table \ref{line_cont_tab} along with the Pearson correlation coefficients. From this diagram, we can see that the range of $F_{2-10}$ is from 2$\times$10$^{-11}$ to 8$\times$10$^{-11}$ (factor of $\sim$4), while that of $I_{\rm FeK\alpha}$ is from 4$\times$10$^{-5}$ to 12$\times$10$^{-5}$ (factor of $\sim$3). Hence, we conclude that \textit{at least} 75$\%$ of the Fe line flux responds to X-ray continuum variations. We searched for lags between $F_{2-10}$ and $I_{\rm FeK\alpha}$ with the interpolated cross correlation function \citep[ICCF;][]{gas87,whi94}, and the discrete cross-correlation function \citep[DCCF;][]{ede88}, with errors determined using the Flux-Randomization and Random Subset Selection (FR-RSS) technique proposed by \citet{pet98}. We omitted the first five points (1997--2003) from this analysis due to the large data gaps present. The ICCF peak correlation coefficient was 0.61 at a continuum-to-line delay $\tau$ = 0 days, with a 1$\sigma$ upper limit to the lag of $\vert \tau \vert = 88 $ days (Figure \ref{line_cont_corfunc}). This implies that the bulk of the line flux originates \textit{within} $\sim$90 light-days of the X-ray continuum source if it lies near the plane of the accretion disk. 
 Figure \ref{line_cont_cor} also demonstrates that $\Gamma_{\rm X}$ increases with $F_{2-10}$, similar to the behavior seen in many Seyfert galaxies that lack a strong jet contribution to the X-ray spectrum \citep{pap02,shi02}. 

\begin{deluxetable}{lcccc}\tablecolumns{5}
\tabletypesize{\footnotesize}
\tablewidth{0pc}
\tablecaption{Best-fit Linear Relations For Parameters Derived From Time-resolved Spectroscopy.\label{line_cont_tab}}
\tablehead{
\colhead{Parameters}             & \colhead{Slope}    & \colhead{y-intercept}   & $r_{\rm P}$ & \colhead{$P_r$}} 
\startdata
$I_{\rm FeK\alpha}$ vs.\ $F_{2-10}$  &  1.27 $\pm$ 0.31 & 1.83 $\pm$ 1.65 & 0.594 & $1.3 \times 10^{-4}$ \\
$\Gamma_{\rm X}$    vs.\ $F_{2-10}$  &  0.040 $\pm$ 0.008 & 1.52 $\pm$ 0.04   & 0.796  & $5.8 \times 10^{-8}$ \\
$EQW$                vs.\ $F_{2-10}$  & $-5.51\pm4.97$ & $150 \pm 29$ &  --1.92   &  0.26   
\enddata
\tablecomments{The parameters for the best-fit linear solutions plotted in Figure \ref{line_cont_cor}. $r_{\rm P}$ is the Pearson correlation coefficient and
$P_r$ is the null hypothesis probability, i.e., the probability of obtaining the observed correlation by chance. \\}
\end{deluxetable}

The X-ray continuum/Fe line correlation presented here is perhaps the strongest such correlation observed so far (cf., for example, the continuum-line correlation observed in NGC 3227 by \citealt{mark09}, where $\sim$50$\%$ of the line flux responded to the continuum variations on time scales of $\lesssim$ 700 days). If the bulk of the Fe line originates within 90 light-days of the X-ray continuum source, then an origin in material commensurate with the optical broad line region (located light-days to light-weeks away from the central black hole in Seyferts) is plausible. However, this delay is inconsistent with a model wherein the bulk of the line emission originates in a $\sim$parsec-scale, homogeneous molecular torus commonly invoked in Seyfert 1/2 unification schemes.\\

\section{Power Spectral Analysis}
\subsection{X-ray}
We use a variant of the Power Spectrum Response method \citep[PSRESP;][]{utt02} to determine the intrinsic PSD of the X-ray light curve. Our realization of PSRESP is described in \citet{cha08}. PSRESP gives both the best-fit PSD model and a ``success fraction'' $F_{\rm succ}$ (fraction of simulated light curves that successfully represent the observed light curve) that indicates the goodness of fit of the model.

\begin{figure}
\epsscale{1.1}
\plotone{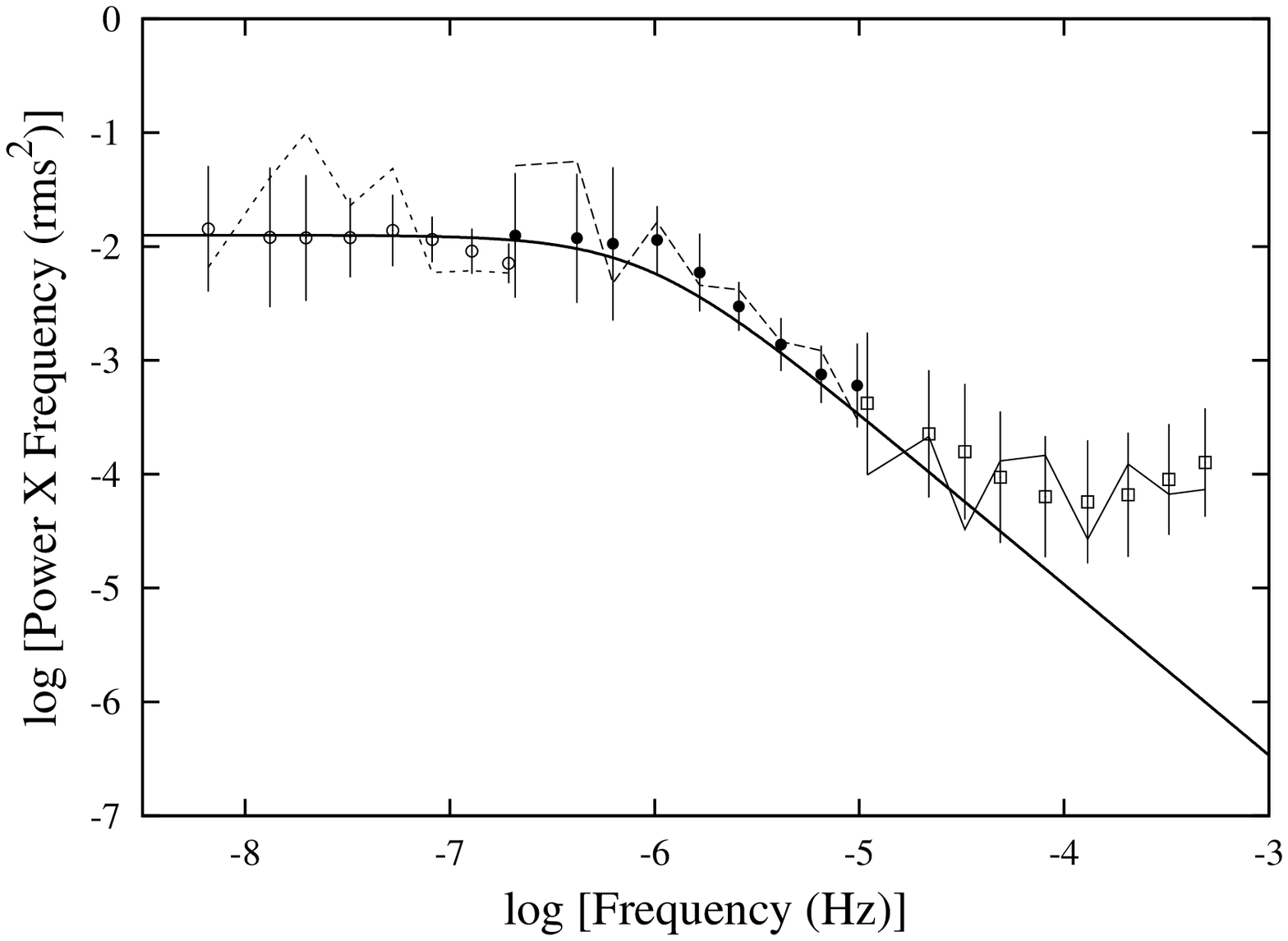}
\caption{Result of application of the PSRESP method to the X-ray light curve of 3C~111. The PSD of the observed data at high, medium, and low frequencies is given by the solid, dashed, and dotted jagged lines, respectively, while the underlying power-law model is given by the thicker solid bent line. Points with error bars (open squares, filled circles, and open circles for high, medium, and low frequency range, respectively) correspond to the mean value of the PSD simulated from the underlying power-law model (see text). The error bars are the standard deviations of the distribution of simulated PSDs. The broadband power spectral density is best described by a bending power law with low frequency slope $-1.0$, high frequency slope $-2.5$, and break frequency $10^{-6.05}$ Hz. The high frequency part of the PSD is dominated by Poisson noise. The figure shows that when the estimated Poisson noise is added to the underlying bending power-law model PSD, it matches the observed PSD quite well. \\ }
\label{psd}
\end{figure}

At first we fit a simple power-law model to the X-ray PSD, but found that the value of $F_{\rm succ}$ is unacceptably low ($0.1$). This implies that a simple power law is not the best model for this PSD. We therefore fit a bending power-law model (broken power law with a smooth break) to the X-ray PSD \citep[see ][]{mch04,cha09},
\begin{equation}
P(\nu)=A\nu^{\alpha_L}[1+(\frac{\nu}{\nu_{B}})^{(\alpha_{L}-\alpha_{H})}]^{-1}.
\end{equation}
Here, $A$ is a normalization constant, $\nu_{B}$ is the break frequency, and $\alpha_H$ and $\alpha_L$ are the slopes of the power laws above and below the break frequency, respectively. During the fitting, we varied the break frequency $\nu_{\rm B}$ from $10^{-9}$ to $10^{-5}$ Hz in steps of $10^{0.05}$, $\alpha_{\rm H}$ (slope of the power law above the break frequency) from $-1.5$ to $-3.0$ in steps of 0.1, and $\alpha_{\rm L}$ (slope below the break frequency) from $-1.0$ to $-1.5$ in steps of 0.1. These ranges include the values of $\alpha$ found in the light curves of BHXRBs, for which $\alpha_{\rm L} \approx -1$ and  $\alpha_{\rm H}$ is between $-2$ and $-3$ \citep[e.g.,][]{rem06}. This procedure yields a much higher success fraction than the simple power-law model. Based on the model with the highest success fraction (0.85), we obtain a good fit with the parameters $\alpha_{\rm L}=-1.0\pm0.1$, $\alpha_{\rm H}=-2.5^{+0.2}_{-0.5}$, and log$_{10}(\nu_{\rm B})=-6.05^{+0.25}_{-0.30}$ Hz, which is equivalent to a timescale $13^{+12}_{-6}$ days. Figure~\ref{psd} presents this best-fit model and the corresponding PSD. As seen in the figure, the high frequency part of the PSD is dominated by Poisson noise. That is because i) this part of the PSD is generated from the longlook light curve, and fluxes in the longlook light curve have larger uncertainties owing to shorter exposure times than those in the other light curves and ii) due to the red noise nature of the PSD, the intrinsic power is smaller at the higher frequencies and hence the power generated by Poisson noise is relatively more important. The figure shows that when the estimated Poisson noise is added to the best-fit model PSD, it matches the observed PSD quite well. 

The PSD break frequency in BHXRBs and Seyfert galaxies scales inversely with the mass of the black hole \citep{utt02,mch04,mch06,ede99,mar03}. \citet{mch06} showed that this holds over a range of BH masses from $10$ to $10^8$ M$_{\sun}$, and that, for a given black hole mass, the break frequency increases with accretion rate ($\dot{m}$). The bolometric luminosity ($L_{\rm bol}$) can serve as a proxy for $\dot{m}$. Since the BH mass in 3C~111 is not well-established, we need to derive its value from independent observations. To do this, we need to determine the optical luminosity of the central engine.

\begin{figure}
\epsscale{1.1}
\plotone{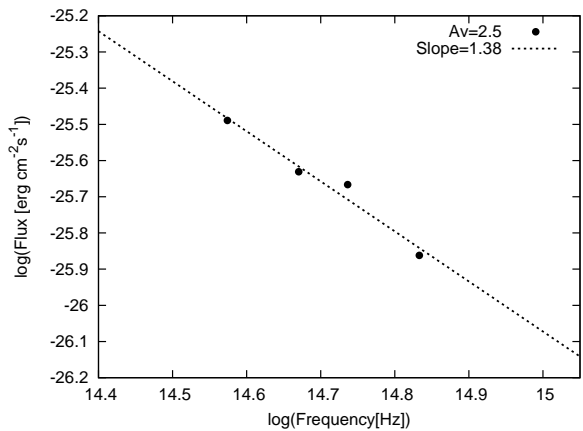}
\caption{The extinction corrected optical spectrum of 3C 111. The best-fit power-law has a slope $-1.38$ for $A_{\rm v}=2.5$. \\}
\label{spec_ind}
\end{figure}

The optical emission received from 3C~111 is subject to substantial extinction owing to the presence of a translucent ($A_{\rm v}$ between $\sim$1 and 5) molecular cloud in the foreground. The standard method for determining the extinction, star counts in the surrounding region on the sky, is subject to systematic uncertainties because of possible gradients in dust absorption. In fact, \citet{MMB93} and \citet{MM95} have reported time variations in radio H$_2$CO absorption lines toward 3C~111, indicating that the column density and/or excitation conditions may vary throughout the cloud, causing changes in absorption as our line of sight to the quasar drifts across the cloud from relative motion between the Earth and the cloud. \citet{sar77} gives essentially simultaneous multi-band optical and near-infrared flux measurements of 3C 111, which are not corrected for extinction. We analyze this spectrum to calculate a value of the extinction, $A_{\rm v}=2.5$, that generates an extinction-corrected spectrum that can be fit well by a power law, the functional form that fits the optical continuum of a typical AGN. The spectral index of the extinction-corrected spectrum, displayed in Figure \ref{spec_ind}, is $-1.38$. Our value of $A_{\rm v}$ is consistent with the values E(B-V) $\sim$1 for translucent clouds with total N(H) $\sim$9$\times10^{21}$ cm$^{-2}$ \citep{rac02} and $A_{\rm v}$/E(B-V) $\sim$2.1 derived for the translucent cloud HD 36982 by \citet{lar96}. We note, however, that these values have high uncertainties and could differ significantly from one cloud to another. After correcting the optical luminosity from \citet{sar77} by this value of $A_{\rm v}$, we arrive at a de-reddened optical luminosity of $\lambda L_{\lambda}=6.1\times 10^{43}$ erg s$^{-1}$ at $\lambda = 510$ nm.

In a recent paper, \citet{dec11} used the properties of the H$\alpha$ line of a sample of blazars in addition to other parameters to determine the mass of the central BH. Using the deprojection factor and H$\alpha$ line to continuum (510 nm) luminosity ratio from equation (3) and (5), respectively, of that paper, as well as the width and luminosity of the H$\alpha$ line from \citet{era03}, we calculate the BH mass of 3C 111, $M_{\rm BH} = 2.4^{+0.6}_{-0.5}\times10^8$ M$_{\sun}$.  Another method for estimating the BH mass is to use the relationship between $M_{\rm BH}$ and the FWHM line width of the H$\beta$ broad emission line along with equation (2) of \citet{ves06}. Unfortunately, no accurate measurement of FWHM(H$\beta$) in 3C 111 has been published; instead, we assume that FWHM(H$\beta$) $\approx$ FWHM(H$\alpha$) = 4800 km~s$^{-1}$ \citep{era03}. From this value and the de-reddened optical luminosity given above, we derive $M_{\rm BH} = 1.5^{+0.4}_{-0.3}\times10^8$ M$_{\sun}$. If we use the more conservative estimate of the uncertainty in the zero point value of the scaling relation between the masses determined using reverberation mapping and those from single-epoch spectra from \citet{ves06}, then the respective uncertainty in the value of the BH mass becomes larger, $M_{\rm BH} = 1.5^{+2.5}_{-1.0}\times10^8$ M$_{\sun}$. Using the correlation between the widths of H$\alpha$ and H$\beta$ emission lines from equation (3) of \citet{gre05}, we calculate FWHM(H$\beta$) of 3C 111 to be $5400\pm400$ km/s from FWHM(H$\alpha$) = 4800 km~s$^{-1}$ \citep{era03}. Using this value in equation (2) of \citet{ves06} and keeping other parameters same as above, we obtain $M_{\rm BH} = 1.8^{+0.5}_{-0.4}\times10^8$ M$_{\sun}$. 

\citet{marche04} derived a much higher value of $M_{\rm BH}$ estimating the bolometric luminosity of 3C 111 from the optical nuclear luminosity along with the bolometric correction from \citet{elv94}. The former is derived by fitting the unresolved nuclear component with the appropriate HST PSF in the respective HST image. They adopted an extinction correction from \citet{SFD98} which is higher than the extinction we calculated above by almost 3 magnitudes. If we adjust the calculation of \citet{marche04} to our extinction correction of $A_{\rm v}$=2.5, we obtain $L_{\rm bol}$ $=(2.5\pm1.0)\times 10^{44}$ ergs s$^{-1}$. We also calculate $L_{\rm bol}$ $=(4.8\pm3.3)\times 10^{44}$ ergs s$^{-1}$ from \citet{you10} using the 2.4--10 keV X-ray flux and 510 nm optical flux, which is consistent with the above value within uncertainties. Performing a similar calculation using equation (11) of \citet{lus10}, we find $L_{\rm bol}$ $=(1.8\pm1.1)\times 10^{45}$ ergs s$^{-1}$, which is close to the above values as well. 

\begin{figure}
\epsscale{1.1}
\plotone{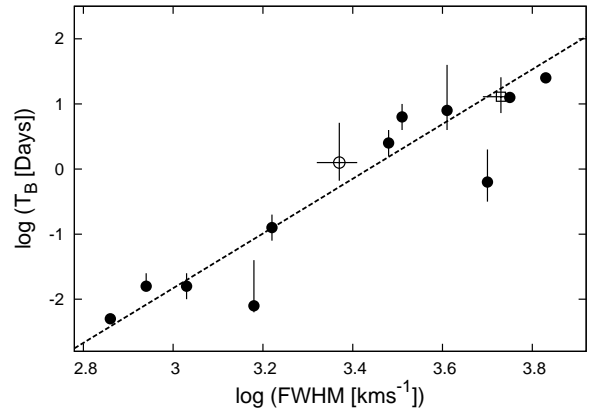}
\caption{The correlation of H$\beta$ line-width with the PSD break timescale for a sample of mainly radio-quiet AGNs from \citet{mch06} is denoted by the dashed straight line and the solid circles. The open circle and the open square show data points corresponding to radio galaxies 3C 120 and 3C 111, respectively, which follow the correlation. For 3C 111, line width of H$\beta$ has been calculated from that of H$\alpha$ using equation (3) of \citet{gre05}. \\}
\label{Tb_Hbeta}
\end{figure}

Using the intermediate values from above, $L_{\rm bol}$ $=(4.8\pm3.3)\times 10^{44}$ erg/s, and $M_{\rm BH} = 1.8^{+0.5}_{-0.4}\times10^8$ M$_{\sun}$, we find Eddington ratio $L_{\rm bol}/L_{\rm edd} = 0.02\pm0.01$. We adopt the best-fit values and uncertainties in the $\rm T_B$---$\rm M_{\rm BH}$---$\rm L_{\rm bol}$ relation proposed by \citet{mch06}, and the above values of $M_{\rm BH}$ and $L_{\rm bol}$. We then calculate the expected break frequency to be $10^{-6.6\pm0.7}$ Hz. If we use the more conservative estimate of the uncertainty in BH mass from above, we obtain $10^{-6.6\pm1.1}$ Hz for the same. Both of these are consistent with the value we have obtained from the best-fit PSD, log$_{10}(\nu_{\rm B})=-6.05^{+0.25}_{-0.30}$ Hz, within uncertainties.
This demonstrates that 3C~111 also follows the $\rm T_B$---$\rm M_{\rm BH}$---$\rm L_{\rm bol}$ relation, the first FR~II radio galaxy for which this has been tested.

\citet{mch06} also showed that FWHM(H$\beta$) in AGNs is strongly correlated with the observed PSD break timescale ($T_B$) given by 
\begin{equation}
\log(T_B)=4.2\log([FWHM(H\beta)])-14.43
\end{equation}
Using the results of the above analysis and that contained in \citet{cha09}, we can now add data points corresponding to the radio galaxies 3C~120 and 3C~111 in Figure 3 of \citet{mch06}, resulting in Figure \ref{Tb_Hbeta}. The plot shows that these two radio galaxies also follow the empirical $T_B$ versus FWHM(H$\beta$) correlation.

\subsection{Optical}
\begin{figure}
\epsscale{1.1}
\plotone{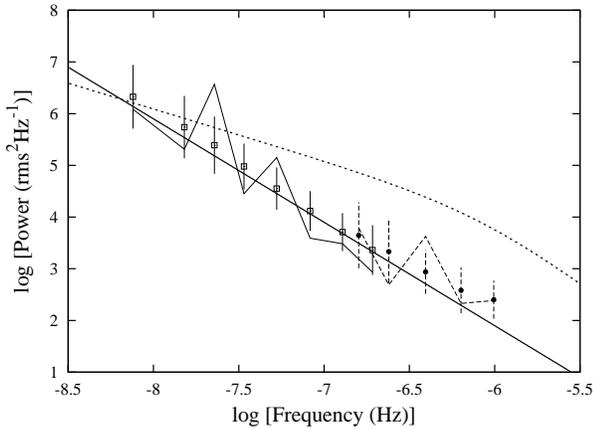}
\caption{Result of application of the PSRESP method to the optical light curve of 3C~111. The PSD of the observed data at high and medium frequencies is given by the dashed and solid jagged lines, respectively, while the underlying power-law model is given by the solid straight line. Points with error bars (filled circles and open squares for high and medium frequency range, respectively) correspond to the mean value of the PSD simulated from the underlying power-law model (see text). The error bars are the standard deviations of the distribution of simulated PSDs. The broadband power spectral density is best described by a simple power law with slope $-2.0$. The smoothly bending dotted line is the underlying best-fit bending power-law model of the X-ray PSD from Fig. \ref{psd}. This shows that the X-ray variability has more power than that in the optical on shorter timescales while on longer timescales variability at the two wave bands tend to be similar. \\}
\label{psd_op}
\end{figure}

We use the same method as above to calculate the PSD of the optical variability. The optical $R$-band PSD of 3C~111 (Figure \ref{psd_op}) shows red noise behavior, i.e., there is higher amplitude variability on longer than on shorter timescales. Based on the model with the highest success fraction, the optical PSD is best fit with a simple power law of slope $-2.0^{+0.3}_{-0.7}$, for which the success fraction is $0.52$. During the fitting, we varied the slope from $-1.0$ to $-3.0$ in steps of 0.1. The rejection confidence, equal to one minus the success fraction, is much less than 0.9, hence the model provides an acceptable fit to the PSD. We also fit a broken power law to the optical PSD, setting the low-frequency slope at $-1.0$ and allowing the break frequency and the slope above the break to vary over a wide range of parameters ($10^{-8}$ to $10^{-6}$ Hz and $-1.0$ to $-2.5$, respectively) while calculating the success fractions. This gives lower success fractions than the simple power-law model across the entire parameter space. Since we do not have any significantly long segment of the $R$-band data when the sampling frequency was more than 2-3 days per week, we determine the optical PSD up to the highest variational frequency that can be achieved with the existing data. A better constraint on the existence of a break in the optical PSD could be achieved with a broader range of sampled frequencies.

The slope of the optical PSD in the range $10^{-8.1}$ to $10^{-5.8}$ Hz is significantly steeper than that in X-rays over the same range ($\sim$$1.0$). Hence, the ratio of shorter to longer timescale variability amplitude is smaller at optical than at X-ray wavelengths. 

\subsection{Excess Variance}
We calculate the fractional root mean squared variability amplitude, i.e., the ``excess variance" normalized by the square of the mean flux, as a measure of the variability in the X-ray and optical data. It is defined as
\begin{equation}
F_{\rm var}=\sqrt{{\frac{\rm S^2-\overline{\sigma_{\rm err}^2}}{\overline{x}^2}}},
\end{equation}
where $\rm S$ is the variance, $\sigma_{\rm err}$ is the observational uncertainty, and $\overline{x}$ is the mean of the data \citep{nan97,ede02,vau03}. We find that the values of $F_{\rm var}$ of the X-ray and optical variability for the entire data set are $0.28$ and $0.14$, respectively. This supports the above conclusion from the PSD analysis that the optical is less strongly variable than the X-ray flux at the longest timescales we probe. It also implies that optical variations cannot be completely responsible for driving all of the X-ray variability observed, if we assume that most of the X-ray emission we observe is not relativistically beamed.\\

\section{X-ray/Radio Correlation}
\begin{figure*}
\epsscale{0.8}
\plotone{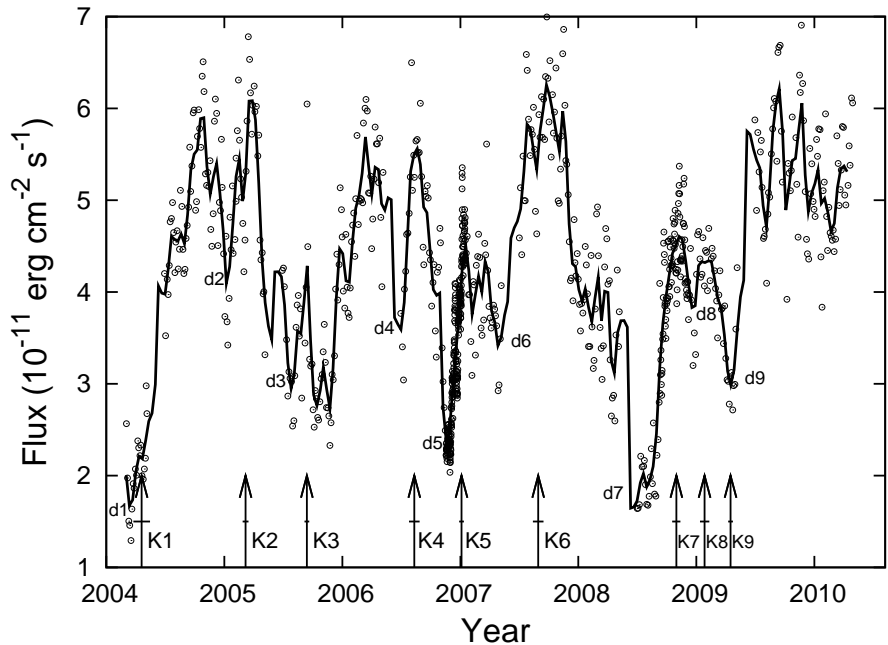}
\caption{X-ray light curve of 3C~111 as in Figure~\ref{xoprad}, with data indicated by small circles. The curve corresponds to the same data, smoothed with a Gaussian function with a 10-day FWHM smoothing time. The arrows indicate the times of superluminal ejections and the line segments perpendicular to the arrows represent the uncertainties in the times. X-ray dips and VLBA knots listed in Table \ref{ejecflare} are marked with the respective ID numbers. Over the six years of observation, significant dips in the X-ray light curve are followed by ejections of bright superluminal knots in the VLBA images. This shows a clear connection between the radiative state near the black hole, where the X-rays are produced, and events in the jet. Dip d3 and the two significant dips just before and after that are considered to be one dip with multiple branches and is related to K3 which was also extended along the jet axis. \\}
\label{xsmooth}
\end{figure*}

We can see in Fig.~\ref{xoprad} that, from the middle of 2009 to the present, the X-ray flux undergoes only minor fluctuations about a mean value of $5\times10^{-11}$ erg cm$^{-2}$ s$^{-1}$. During this interval, the 230 GHz flux is at a very low level, while the 15 GHz flux is decreasing. Therefore, the high-state X-ray plateau coincides with a low state of the radio jet. We can extend this X-ray/radio connection to the observations of earlier years as well. The times of ``ejection'' of new radio knots in the jet (shown by the arrows) are related to the low X-ray states. We convolve the X-ray light curve with a Gaussian smoothing function with a 10-day FWHM smoothing time to identify the major long-term trends. Figure \ref{xsmooth} shows the smoothed X-ray light curve and denotes the times of superluminal ejections with arrows. Based on inspection of this light curve, we consider an X-ray fluctuation to be a significant dip if the smoothed X-ray flux falls by more than 30\% and remains at a low level for longer than one month. We calculate the central time of each dip by determining the local minimum of the X-ray flux. We note that the minimum amplitude measured from unsmoothed data of every dip is below the average of the X-ray flux over the length of the light curve ($4\times10^{-11}$ erg cm$^{-2}$ sec$^{-1}$). It can be seen that each of the ejections is preceded by a significant dip in the X-ray flux. The prolonged low level of X-ray flux in 2005 is associated with knot K3, which was extended along the jet axis, as is apparent in the images starting in late 2006 (Fig. \ref{vlbaimages}). This suggests that a prolonged disturbance near the base of the jet resulted in an elongated superluminal knot. Hence, d3 and the two significant dips just before and after that are considered to be one dip with multiple branches and is related to K3. Detection of the ejection of a new knot requires the analysis of a sequence of VLBA images of sufficient duration to define the trajectory and speed of a knot, as is evident from Figures \ref{vlbaimages} and \ref{distepoch}. No new bright, moving knots after K9 appeared in subsequent imaging with the VLBA during the first 8 months of 2010 (images not shown here).

\begin{figure}
\epsscale{1.1}
\plotone{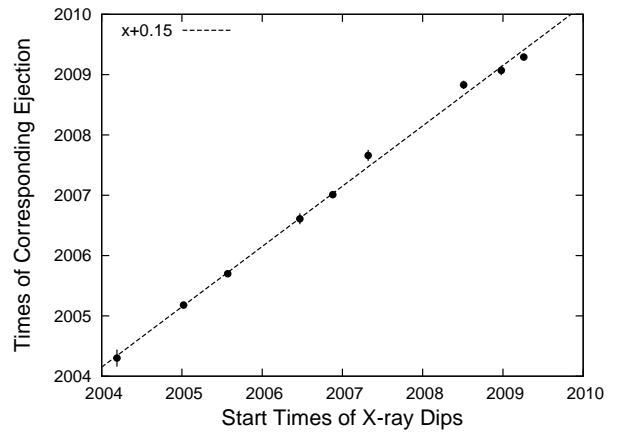}
\caption{Times of ejection of new VLBA knots versus times of X-ray dips from Table \ref{ejecflare}. The dashed line is the best-fit straight line through the points. It fits the data extremely well with small scatter. This indicates that there is a clear association between X-ray dips and superluminal ejections. The y-intercept of this line minus 2004.0 indicates the value of mean time delay between the times of dips and ejections. \\}
\label{dipejec_stline}
\end{figure}

The time delay between the minimum of the X-ray dips and the time of ejection of the corresponding superluminal knot is distributed between $0.03$ and $0.34$ yr (Table \ref {ejecflare}) with a mean of $0.15\pm0.08$ yr. We plot the times of ejection of new knots along with the corresponding times of X-ray dips in Figure \ref{dipejec_stline}. A straight line fits the data extremely well with small scatter, which indicates that there is a clear association between X-ray dips and superluminal ejections. The best-fit line through the points has a slope of 1 and y-intercept of $0.15$, which is identical to the mean delay of $0.15\pm0.08$ yr given above. We have performed a numerical simulation to calculate the significance of the dip-ejection correlation. If we keep the times of dips fixed and choose one million sets of nine ejections, times of which are drawn from a uniform random distribution between 2004.3 and 2010.3 (the duration of our observations), the probability that there is at least one ejection event following all 9 dips by $0.03$ to $0.34$ yr is $0.007$. In a similar simulation, we found that the probability that all 9 dips will have exactly one following ejection is $3\times10^{-5}$. This shows that the dip-ejection association in 3C~111 is highly significant. To further investigate the relationship between the characteristics of the X-ray dips and VLBA knots, we calculate the average flux over all epochs of the latter. We do this via modeling of the brightness distribution at each epoch with multiple components characterized by circular Gaussian brightness distributions, as described in {\S}2.4. This shows that the average flux of the VLBA knots corresponding to the shallower dips (d2, d4, and d8 in Fig. \ref{xsmooth}) is significantly smaller ($0.3\pm0.1$ Jy) than that for the larger dips (d1, d3, d5, and d7), $0.9\pm0.4$ Jy. Dip d6 is also shallow, but this may be due to the absence of data during the Sun-avoidance gap, while d9 is of intermediate depth. Hence K6 and K9 were not included in the average calculation. That the brighter knots correspond to more pronounced dips supports the conclusion that the X-ray minima and the ejection of knots in the jet are physically related.

\begin{figure}
\epsscale{1.1}
\plotone{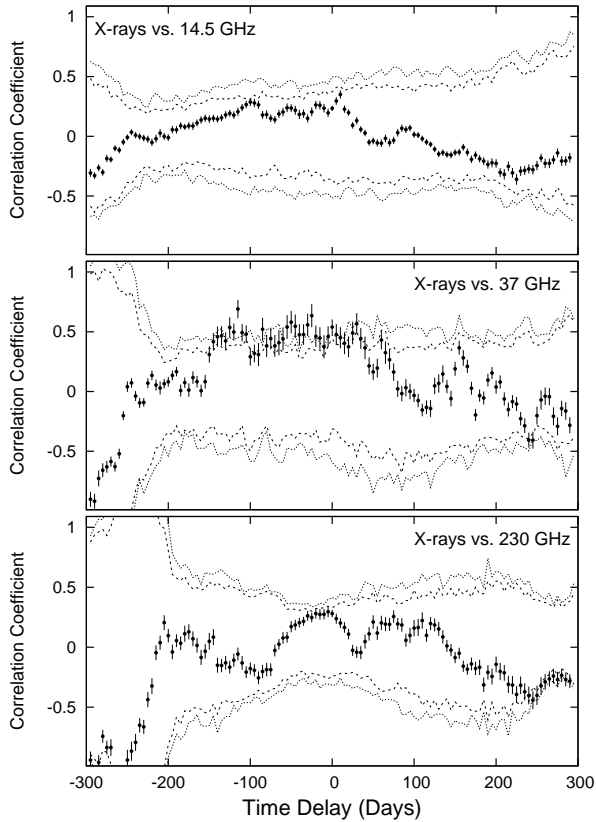}
\caption{Discrete cross-correlation function (DCCF) of the X-ray and radio monitoring data of 3C~111. The time delay is defined as positive if the X-ray variations lag those at radio frequencies. Top panel shows the correlation function for X-ray versus 15 GHz variations while the middle and bottom panels show the same for X-ray versus 37 and 230 GHz, respectively. The jagged dotted and dashed lines at the top of each panel denote 99 and 95 percent significance levels of correlation, respectively (see text). Those at the bottom of each panel show the same for anti-correlation. This shows that the X-ray and 37 GHz variations are correlated while the other two pairs are not significantly correlated. \\}
\label{xrad}
\end{figure}

As determined by the discrete cross-correlation function \citep[DCCF;][]{ede88} shown in the top panel of Figure~\ref{xrad}, the X-ray flux variations are weakly correlated with those at 15 GHz in 3C~111. The peak X-ray versus 15 GHz DCCF is $0.35$. We simulate X-ray light curves from the underlying PSD as determined in {\S}3 and resample the simulated light curves with the sampling window of the actual X-ray light curve. Then we cross-correlate the resampled simulated X-ray light curves with the real 15 GHz light curve. The solid and dotted jagged lines in Figure~\ref{xrad} indicate 99\% and 95\% extremes of the distribution of correlation coefficients as a function of time delay. 

The DCCF indicates that the X-ray flux variations are correlated with those at 37 GHz in 3C~111, with a peak value of $0.63$ (middle panel of Figure~\ref{xrad}). The position of the peak of the correlation function indicates the relative time delay between the variations at the two wavelengths. The time lag of the peak corresponds to X-ray variations leading those at 37 GHz by $40^{+70}_{-65}$ days. We use the FR-RSS technique \citep{pet98} to calculate the mean value and the uncertainty of the cross-correlation time lag. The bottom panel of Figure~\ref{xrad} shows that the X-ray and 230 GHz variations are weakly correlated, with a peak DCCF of $0.30$. 

We note that if the radio flux increases within tens of days after an X-ray minimum and the X-ray flux recovers from the minimum to increase as well during this time, then a relatively weak X-ray/radio correlation will result as seen in the data presented in this paper. The significant correlation between the dips and ejections of new knots in the jet implies that increased activity in the radio jet of 3C~111 is related to a temporary {\it decrease} in X-ray production. Similarly, the association between the high-state X-ray plateau with a low state of the radio jet after the middle of 2009 suggests that a persistently high X-ray flux is associated with a low radio state.

It is possible that the ``corona,'' where the X-ray emission seen in AGN is thought to arise from Compton up-scattering of softer accretion-disk photons, might be the base of the jet \citep{mar05}. If this is the case, then the X-ray flux will be related to the number of electrons residing there and available for scattering to create X-rays. If the same number of electrons is injected into the base of the jet per unit time, then faster flow velocities will correspond to lower densities in the scattering region. The mass loading of the jet should also affect the asymptotic Lorentz factor of the flow downstream if the jet is magnetically driven \citep[e.g.,][]{vla04}. The same decrease in electron number that causes a drop in scattered X-ray emission near the disk would lead to a time-delayed increase in the speed of the jet downstream. The increase in flow speed of the jet could form a shock wave, eventually seen as a superluminal radio knot \citep{gom97}. This would then give rise to the dip-ejection sequence discussed above.

The highest-amplitude 37 GHz outburst, from 2007.5 to 2008.8, started at the same time as at 230 GHz, but reached maximum level at a later time than both the 230 GHz and X-ray flares. This can be explained by the larger optical depth at the lower frequencies prior to the 37 GHz peak. The similar amplitudes of the 37 and 230 GHz flares follow the pattern of the synchrotron-loss stage of the shock-in-jet model \citep{mar85}, which predicts that, in synchrotron flares produced in the jet, the peak amplitude should stay roughly constant as the emission becomes optically thin at progressively lower frequencies.

The mean time delay between the minimum of the X-ray dips and the time of ejection of the corresponding superluminal knot is $0.15\pm0.08$ yr. The average apparent speed of the moving components with well-determined motions is $3.9c\pm0.7c$. Therefore, a knot moves a distance of $0.18\pm0.1$ pc in $0.15$ yr, projected on the plane of the sky. (Here we assume that acceleration of the flow to its terminal velocity occurs over a sufficiently short distance that the time that a moving feature spends in the acceleration zone is short compared with the total transit time to the 43 GHz core.) Since the angle of the jet axis of 3C~111 to the line of sight $\sim$18$^{\circ}$, the actual distance traveled by the knot, given by $\beta c \delta \Delta t_{\rm obs} \Gamma$ is $0.6\pm0.3$ pc. Hence, we derive a distance $0.6\pm0.3$ pc from the corona or the base of the jet (where the X-rays are produced) to the core seen on the 43 GHz VLBA images. This is one of the few cases where, similar to 3C~120 \citep{cha09}, we are able to specify the distance between the central engine and mm-wave core in an AGN.\\ \\ \\

\section{X-ray/Optical Correlation}
\begin{figure}
\epsscale{1.2}
\plotone{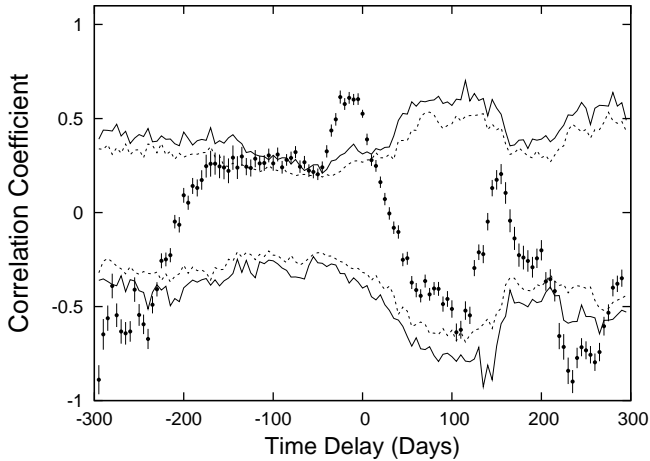}
\caption{Discrete cross-correlation function (DCCF) of the optical and X-ray monitor data of 3C~111 for the entire 5 yr interval. The time delay is defined as positive if the X-ray variations lag those at the optical frequency. The jagged solid and dotted lines at the top of the panel denote 99 and 95 percent significance levels of correlation (see text), respectively. Those at the bottom of the panel show the same for anti-correlation. The peak X-ray versus optical DCCF is $0.6$, which corresponds to more than 99\% significance level. The peak at $-15$ days indicates that the X-ray variations lead those in the optical by $15\pm10$ days. \\}
\label{xop}
\end{figure}

As determined by the DCCF (Figure~\ref{xop}), we find that the X-ray variations of 3C~111 are very strongly correlated with those at optical $R$ band. The peak X-ray versus optical DCCF is $0.6$, which corresponds to more than 99\% significance level. The jagged solid and dotted lines at the top of the panel denote 99 and 95 percent significance levels of correlation, respectively. Those at the bottom of the panel show the same for anti-correlation, as described in the previous section. The time lag of the peak indicates that the X-ray variations lead those in the optical by $15\pm10$ days. The highly significant correlation between the X-ray and optical variations implies that emission at these wave bands is causally connected. 

Fig.~\ref{op_pol} shows the variation of the degree of polarization as well as the electric vector position angle (EVPA) of 3C~111 at $R$ band from 2006 to the present. The average polarization was only $1.6\%\pm0.6\%$, rather low for synchrotron radiation from the jet. This agrees with the idea that most of the optical emission is thermal radiation from the accretion disk \citep{mal83}. The strong X-ray/optical correlation, weak optical polarization, and smaller variance of optical compared with X-ray flux at shorter timescales are consistent with a reprocessing model. In this scenario, the X-rays are predominantly produced by inverse Compton (IC) scattering of thermal optical/UV seed photons from the accretion disk by hot, but non-relativistic, electrons in the corona, while a significant fraction of the optical-UV emission is due to heating of the accretion disk by X-rays produced in the above process. The slope of the optical PSD in the range $10^{-8.1}$ to $10^{-5.8}$ Hz is $\sim$$-2.0$, significantly steeper than that of the X-ray PSD over the same frequency range ($\sim$ $-1.0$). This implies that the optical variations on shorter time-scales are suppressed, consistent with smoothing by reprocessing. 

Alternatively, a substantial fraction of the optical emission may be generated as synchrotron radiation in the jet, giving rise to a degree of polarization up to 3\% (Fig. \ref{op_pol}) or even higher \citep{jor07} and significant variability of the position angle of polarization (between $0^\circ$ and $150^\circ$) at some epochs. In this case, if X-rays are indeed generated at the base of the jet, then the highly significant correlation between the X-ray and optical variations, combined with the time delay of the latter, $15\pm10$ days, implies that the optical emission region in the jet, if present, is situated downstream of the base of the jet. We can calculate the mean distance $(\Delta z)_{\rm xo}$ between the X-ray and optical synchrotron emission sites in the jet if we adopt $18^\circ$ as the angle between the jet axis and the line of sight \citep{jor05} and assume a value for the average speed $\beta c$ of the jet flow between the two emission regions: $(\Delta z)_{\rm xo} \sim $ 6, 20, and 200 light-days for $\beta$ = 0.25, 0.5, and 0.974, respectively. The last value, $\beta = 0.974$, corresponds to the velocity obtained from analysis of the superluminal apparent motion and time-scale of flux decline at 43 GHz of knots observed with the VLBA \citep{jor05}. The lower values of $\beta$ near the base of the jet are in concert with theoretical models in which jets are accelerated to relativistic flow velocities over extended distances from the central engine \citep[e.g.,][]{vla04}.

Some of the thermal optical-UV emission could be produced in the accretion disk even without reprocessing via X-ray heating. The temporal nature of this emission would be different from that generated as a result of reprocessing. The X-rays may be produced mainly by up-scattering of UV rather than optical photons. This could occur if the flux of optical photons reaching the corona is much smaller than that of UV photons. Such a scenario is likely if the corona is small, such that the UV emission region is much closer to the corona than the region where the majority of the optical photons are produced \citep[see][for details]{cha09}. Any disturbance propagating outward in the accretion disk will cause a change in the UV flux (and a resultant nearly immediate change in the X-ray emission) followed by a similar change in the optical flux. This may also give rise to the X-ray/optical time delay that is observed. As discussed in detail by \citet{cha09}, the variability timescale described above is consistent with the model proposed by \citet{kin04} and \citet{liv03}, in which variability in the disk emission is caused by large-scale alignment of poloidal magnetic field in the inner accretion disk from random fluctuations in field direction. Such alignment occurs on a timescale of a few tens of days for a BH of mass of $\sim$$10^8$M$_{\sun}$, consistent with the results described above.

The time delay of $15\pm10$ days of the optical with respect to the X-ray variations is larger than expected from a pure reprocessing model. As discussed in \citet{cha09}, the region of the accretion disk that produces the largest amount of direct (rather than reprocessed) optical emission is at $\sim$100 gravitational radii ($r_g$) from the center, which is equal to $\sim$1 light-day for a BH mass of $1.8\times10^8$ M$_{\sun}$ \citep[see also the detailed modeling of][]{kaz01}. Therefore, if the X-ray/optical time delay is solely due to light travel time from the corona to the region in the accretion disk producing reprocessed optical emission, the corresponding reprocessing region has to be farther and/or extended than this by an order of magnitude ($\sim$1000 $r_g$). If, on the other hand, the variability is caused by disturbances propagating outward through the disk at velocity v$_{\rm dist}$, then the optical emission region could be closer by a factor v$_{\rm dist}$/c. If a fraction of the optical emission is produced in the jet, as discussed above, 
or the variability is partly affected by disturbances propagating in the accretion disk, then the observed time delay and flat peak of the X-ray/optical cross-correlation function could result. 

The accretion rate estimated in {\S}3, $\dot{m}=(0.02\pm0.01)\dot{m}_{\rm edd}$, is within the range at which the transition from high to low state occurs in the Galactic BHXRBs \citep{mac03}. An advection dominated accretion flow (ADAF) solution may exist for accretion onto black holes at this rate \citep{nar05}. The radiation in an ADAF is dominated by thermal comptonization, which has been used to explain the X-ray spectra of BHXRBs in the canonical low-hard state \citep[e.g.,][]{esi98}. It is possible that the X-rays from 3C 111 that we have attributed to a distribution of hot electrons consistent with the corona might instead be produced by an ADAF.\\

\section{Comparison with the FR~I Radio Galaxy 3C~120}
Although 3C~111 and 3C~120 fall into two different Fanaroff-Riley categories, they have a number of common characteristics. Both are radio-loud AGNs with jets that dominate the emission at radio frequencies, but with less radiative power than typical blazars and directed at a wider angle to our line of sight. In both radio galaxies, the low optical polarization is consistent with most of the optical continuum resulting from blackbody radiation from the accretion disk. The X-ray properties are consistent with inverse Compton scattering of thermal optical-UV photons from the disk by a distribution of hot electrons --- the corona, which could be the base of the jet --- situated near the disk. In 3C~111, some of the optical and X-ray emission could be from the jet. A major difference between these two radio galaxies is in the accretion rate. In 3C 120, the accretion rate is high ($\sim$0.3 $\dot{m}_{\rm edd}$), similar to Galactic BHXRBs in the X-ray high state. In 3C 111, it is lower by an order of magnitude ($\sim$0.02 $\dot{m}_{\rm edd}$) which is close to the transition from high to low state in galactic BHXRBs. Therefore, the inner accretion flow in 3C 111 may be advection dominated.

The X-ray emission from each of these radio galaxies possesses a characteristic timescale that is proportional to the BH mass, similar to what is observed in Galactic BHXRBs and Seyfert galaxies. This result implies that there is a universality in the accretion processes of BHs in the mass range $10 M_{\sun}$ to $10^8 M_{\sun}$. These objects also show a clear connection between the radiative state near the BH and events in the jet. Ejection of new knots in the jet following dips in the X-ray light curve observed in these two radio galaxies is somewhat similar to the Galactic micro-quasar GRS 1915+105, where knots with apparent superluminal motion appear in the jet after the X-ray flux recovers from intervals of very low flux levels \citep{mir98}. This connection is expected according to current theories of jet launching and collimation, but has not been demonstrated previously to extend to radio-loud AGNs.

The relationships that we have uncovered between the central engine and the jet provide strong support for the paradigm \cite[e.g.,][]{kor06, marecki11} that AGNs and Galactic BHXRBs are fundamentally similar, with characteristic time and size scales proportional to the mass of the central BH. This implies that we can develop and test models of AGNs based partly on observations of BHXRBs.\\

\section{Summary and Conclusions}
This paper presents well-sampled, 6-yr-long light curves of 3C~111 between 2004 and 2010 at X-ray, optical, and radio wavebands, as well as monthly images obtained with the VLBA at 43 GHz. The appearance of new knots in the VLBA images of the jet follow prominent dips in the smoothed X-ray light curve, demonstrating a connection between events in the jet at parsec scales and those occurring near or in the accretion disk. We have calculated the broad-band X-ray power spectral density (PSD) of 3C~111 and cross-correlated the X-ray and optical light curves, determining the significance of the correlations with light curves simulated in accordance with the PSD. From the analysis of the temporal properties of the observed variability in multiple wave bands, we infer the locations and radiative mechanisms of the X-ray and optical emission.  

Our main conclusions are as follows: \\
(1) The X-ray PSD of 3C~111 is best fit by a bending power-law model with a smooth change in slope above a break frequency. The best-fit value of the break frequency is $10^{-6.05}$ Hz, corresponding to a break timescale of $13$ days, which agrees with the empirical relation between break timescale, BH mass, and accretion rate obtained by \citet{mch06} based on a sample of mainly radio-quiet Seyfert galaxies. This indicates that the accretion process in 3C~111 is similar to that of BHXRBs and Seyfert galaxies, and that there is a possible universality in the accretion processes in BHs spanning a broad range of masses from 10 $M_{\sun}$ to $10^8$ $M_{\sun}$.\\
(2) In 3C~111, the 2--10 keV continuum flux and the Fe line intensity are strongly correlated, with a time lag shorter than 90 days and consistent with zero. This implies that the Fe line is generated within 90 light-days of the source of the X-ray continuum.\\
(3) All X-ray dips are followed by the ejection of a new knot in the VLBA images. We derive a distance from the base of the jet or corona (where the X-rays are produced) to the VLBA 43 GHz core region of $\sim$0.6 pc using the average time delay between the minimum of the X-ray dips and the time of passage of the corresponding superluminal knots through the core. \\
(4) The dip-ejection connection can be explained if a decrease in the X-ray production is linked with increased speed in the jet flow, causing a shock front to eventually form and move downstream. This property of 3C~111 is similar to 3C~120 and reminiscent of Galactic BH systems, where transitions to high-soft X-ray states are associated with the emergence of very bright features that proceed to propagate down the radio jet.\\
(5) Most of the optical emission in 3C~111 is generated in the accretion disk due to X-ray heating. However, weak optical polarization indicates that some fraction of the optical emission is synchrotron radiation from the jet. \\

\section{Acknowledgments}
The research at Boston University (BU) was funded in part by the National Science Foundation (NSF) through grants AST-0406865 and AST-0907893 and by NASA through Astrophysical Data Analysis Program grant NNX08AJ64G and XMM-Newton grant NNX08AY06G. The PRISM camera at Lowell Observatory was developed by K. Janes et al. at BU and Lowell Observatory, with funding from the NSF, BU, and Lowell Observatory. The University of Michigan Radio Astronomy Observatory is supported by funds from the NSF, NASA, and the University of Michigan. The VLBA is an instrument of the National Radio Astronomy Observatory, a facility of the National Science Foundation operated under cooperative agreement by Associated Universities, Inc. The Liverpool Telescope is operated on the island of La Palma by Liverpool John Moores University in the Spanish Observatorio del Roque de los Muchachos of the Instituto de Astrofisica de Canarias, with financial support from the UK Science and Technology Facilities Council. The Mets\"ahovi team acknowledges the support from the Academy of Finland. The research at IAA-CSIC is supported by the Spanish Ministry of Science and Innovation grants AYA2007-67627- C03-03 and AYA2010-14844, and by the Regional Government of Andaluc\'{\i}a (Spain) grant P09-FQM-4784. This paper is partly based on observations carried out at the German-Spanish Calar Alto Observatory, which is jointly operated by the MPIA and the IAA-CSIC. Acquisition of the MAPCAT data is supported in part by MICIIN (Spain) grants AYA2007-67627-C03-03 and AYA2010-14844, and by CEIC (Andaluc\'{i}a) grant P09-FQM-4784. The Submillimeter Array is a joint project between the Smithsonian Astrophysical Observatory and the Academia Sinica Institute of Astronomy and Astrophysics and is funded by the Smithsonian Institution and the Academia Sinica.\\

\end{document}